%% file: ms.tex
\documentclass{emulateapj}

\usepackage{subfigure}
\usepackage{url}
\usepackage{hyperref}
\usepackage{datetime}
\usepackage{longtable}
\usepackage{natbib}
\usepackage{amsmath}
\usepackage{ulem}
\usepackage{bm}
\usepackage[usenames,dvipsnames]{xcolor}
\usepackage[T1]{fontenc}

\newcommand{\noprint}[1]{}

\usepackage{color}

\newcommand{\ars}{{$a/R_{\star}$}}

\newcommand{\project}[1]{\textsl{#1}}
\newcommand{\kep}{\project{Kepler}}
\newcommand{\KT}{\project{K2}}
\newcommand{\Ci}{Campaign~1}

\newcommand\teff{\ensuremath{T_\text{eff}}}

\newcommand{\rsun}{{R$_\odot$}}
\newcommand{\msun}{{M$_\odot$}}

\newcommand{\feh}{{[Fe/H]}~}
\newcommand{\afe}{{[$\alpha$/Fe]}~}

\newcommand{\paperit}{\citet{Foreman-Mackey15}}

\newcommand{\Nfp}{6}
\newcommand{\Nvalidated}{21}
\newcommand{\Nvalnew}{17} 

\begin{document}

\title{Stellar and Planetary Properties of \KT\ Campaign 1
Candidates and Validation of \Nvalnew\ Planets, Including a Planet Receiving
Earth-like Insolation}

\newcommand{\caltech}{1}
\newcommand{\cfa}{2}
\newcommand{\nyu}{4}
\newcommand{\cds}{5}
\newcommand{\princeton}{3}
\newcommand{\mpia}{6}
\newcommand{\jcpa}{7}
\newcommand{\texas}{8}
\newcommand{\hjs}{9}

\author{%
Benjamin~T.~Montet\altaffilmark{\caltech,\cfa},
Timothy~D.~Morton\altaffilmark{\princeton},
Daniel~Foreman-Mackey\altaffilmark{\nyu,\cds},
John~Asher~Johnson\altaffilmark{\cfa},
David~W.~Hogg\altaffilmark{\nyu,\cds,\mpia},
Brendan~P.~Bowler\altaffilmark{\caltech,\jcpa},
David~W.~Latham\altaffilmark{\cfa},
Allyson~Bieryla\altaffilmark{\cfa},
Andrew~W.~Mann\altaffilmark{\texas,\hjs},
}

\email{btm@astro.caltech.edu}
\altaffiltext{\caltech}  {Cahill Center for Astronomy and Astrophysics,
                          California Institute of Technology, Pasadena, CA,
                          91125, USA}
\altaffiltext{\cfa}      {Harvard-Smithsonian Center for Astrophysics,
                          Cambridge, MA 02138, USA}
\altaffiltext{\princeton}{Department of Astrophysics, Princeton University,
                          Princeton, NJ, 08544, USA}
\altaffiltext{\nyu}      {Center for Cosmology and Particle Physics,
                          Department of Physics, New York University,
                          4 Washington Place, New York, NY, 10003, USA}
\altaffiltext{\cds}      {Center for Data Science,
                          New York University,
                          726 Broadway, 7th Floor, New York, NY, 10003, USA}
\altaffiltext{\mpia}     {Max-Planck-Institut f\"ur Astronomie,
                          K\"onigstuhl 17, D-69117 Heidelberg, Germany}
\altaffiltext{\jcpa} {Caltech Joint Center for Planetary Astronomy Fellow}
\altaffiltext{\texas} {Department of Astronomy, The University of Texas at
                            Austin, Austin, TX 78712, USA}
\altaffiltext{\hjs} {Harlan J. Smith Fellow, The University of Texas at Austin}

\date{\today, \currenttime}

\begin{abstract}
The extended \kep\ mission, \KT, is now providing photometry of new fields every
three months in a search for transiting planets.
In a recent study, Foreman-Mackey and collaborators presented a list of 36 planet
candidates orbiting 31 stars in \KT\ \Ci.
In this contribution, we present stellar and planetary properties for all systems.
We combine ground-based seeing-limited survey data and
adaptive optics imaging with an automated transit analysis scheme to validate
\Nvalidated\ candidates as planets, \Nvalnew\ for the first time,
and identify \Nfp\ candidates as likely false positives.
Of particular interest is K2-18 (EPIC 201912552), a bright ($K=$8.9) M2.8 dwarf
hosting a $2.23 \pm 0.25$ $R_\oplus$ planet with $T_{eq} = 272 \pm 15$ K and an
orbital period of 33 days.
We also present two new open-source software packages which enable this analysis.
The first, \texttt{isochrones}, is a flexible tool for fitting theoretical stellar
models to observational data to determine stellar properties using a nested sampling
scheme to capture the multimodal nature of the posterior distributions of the physical 
parameters of stars that may plausibly be evolved.
The second is \texttt{vespa}, a new general-purpose procedure to calculate false 
positive probabilities and statistically validate transiting exoplanets.
\end{abstract}


\maketitle

\section{Introduction}
The \kep\ telescope \citep{Borucki10} has led to a revolution in stellar and planetary
astrophysics, with 7305 ``objects of interest'' and 4173 ``planet candidates''
discovered to date \citep{Borucki11a, Borucki11b, Batalha13, Burke14, Mullaly15, Rowe15}.
The fidelity of this sample is high: most of these candidates are truly planets
\citep{Morton11b, Fressin13, Desert15}.
The mechanical failure of two reaction wheels on the spacecraft led to a repurposing
of the spacecraft into the \KT\ mission, in which the telescope points at
fields near the ecliptic plane for $\sim 75$ days at a time \citep{Howell14}.
In this observing strategy, two axes of motion of the spacecraft are
controlled by the two remaining reaction wheels, while the roll of the
spacecraft is balanced with Solar radiation pressure and quasiperiodic
thruster firing.
As a result, the detector drifts relative to the sky
at the rate of $\sim 1\arcsec$ hr$^{-1}$, with rapid corrections due to thruster fires
approximately once every six hours.
Over the full duration of each campaign, the targets remain near the
same location on the detector but both the slow drift and the corrections are
observable by eye \citep{Barentsen15}.

\KT\ light curves produced with aperture photometry contain substantial pointing-induced
photometric variations caused by the star's apparent motion over a poorly defined flat
field.
Worse yet, these variations occur on timescales similar to transit signals, potentially
masking the observational signature of a planet passing between \kep\ and its host star.

There has been considerable effort to recover these planetary signals, and to date six
planets have been confirmed orbiting three stars in the \KT\ data
\citep{Armstrong15b, Crossfield15, Vanderburg15}.
What is common to all of these methods are that removal of systematics is considered a
step to be undertaken before the search for planets.
Under this strategy, it is implicitly assumed that the systematics are removed perfectly,
while retaining all of the astrophysical signal.
Of course, it is impossible to perfectly separate the astrophysical and instrumental
signal, and such a technique is prone to either over-fitting, in which some of the
astrophysical signal is also removed, or under-fitting, in which some
of the instrumental systematics remain.
A better strategy is to simultaneously fit both the signal and the systematics, as is common
practice in cosmology and, increasingly, in radial velocity searches for planetary systems
\citep[e.g.][]{Ferreira00, Boisse11, Haywood14, Grunblatt15}.

\paperit{} simultaneously fit both the systematics and potential planetary transit signals
in a search for transiting planets.
They assume that the dominant trends in the observed stellar light
curves are caused by spacecraft motion and are shared by many stars.
They then run principal component analysis (PCA) on all stars to measure the dominant modes, 
modeling each star as a
linear combination of 150 of these ``eigen light curves'' and a transit signal.
This method enables fitting without over-fitting, and also permits marginalization over
uncertainties induced by the systematic model.
Therefore, any uncertainties in the systematics can be propagated into uncertainties in
detected planet parameters, instead of assuming the systematics are understood perfectly.
Using this technique, \paperit{} detect 36 planet candidates orbiting 31 stars in \KT\
\Ci\ data.

In \paperit, only transit properties are provided, not absolute parameters about the
planet or the star.
Additionally, the authors follow the convention of the \kep\ team to include any
transit event as a candidate system rather than a false positive if a secondary eclipse
is not detected: there is no enforced upper limit on the allowed planet radius.
The authors intentionally make no effort to separate true transiting planets from
astrophysical events that mimic the appearance of transits, such as an eclipsing
binary (EB) with a high mass ratio, similar to the \kep\ team's list of ``objects of interest.''

In this paper, we present stellar and planetary parameters for each system.
We also analyze the false positive probability (FPP) of each system using \texttt{vespa},
a new publicly available, general-purpose implementation
of the \citet{Morton12} procedure
to calculate FPPs for transiting planets.
Through this analysis, as well as archival imaging, ground-based
seeing-limited survey data, and adaptive optics imaging, we are able to confirm
\Nvalidated\ of these systems as transiting planets at the 99\% confidence
level.
Additionally, we identify six systems as false positives.

This paper is organized as follows.
In Section 2, we develop stellar properties through
photometric and spectroscopic data.
In Section 3, we combine the derived stellar properties with \KT\ data to infer planet
candidate properties.
In Section 4, we combine adaptive optics and radial velocity observations with both
archival and modern ground-based, seeing limited survey data and an analysis of the
transit parameters to calculate FPPs.
In Section 5, we discuss potentially interesting systems, including a mini-Neptune
orbiting an M dwarf which receives a similar insolation to the Earth.
In Section 6, we summarize and discuss our results.

\section{Stellar Properties}
\subsection{Photometry}

\begin{figure}[htbp]
\centerline{\includegraphics[width=0.5\textwidth]{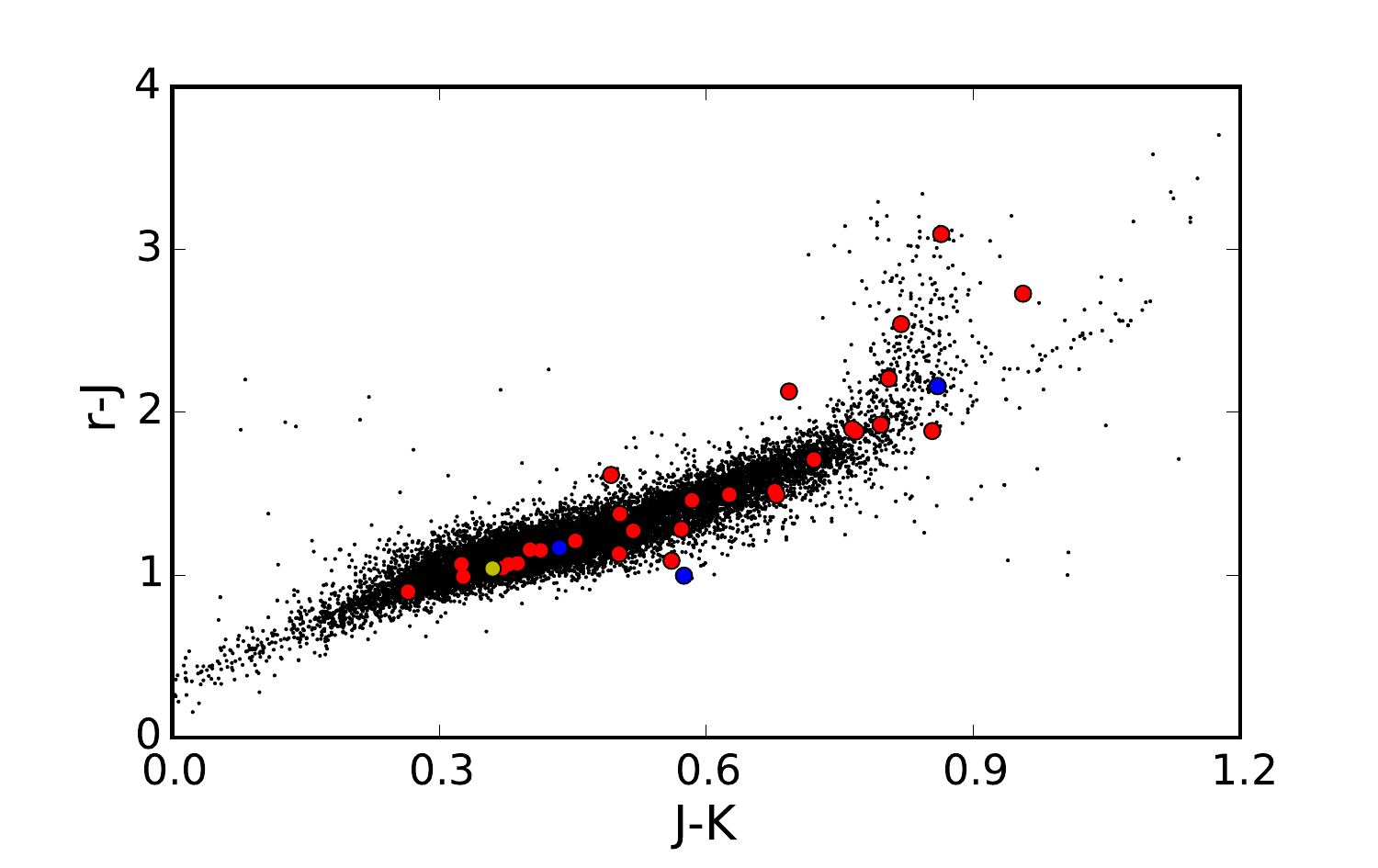}}
\caption{Color-color diagram displaying $r-J, J-K$ photometry for targets
observed by \kep\ during the original
mission (black), with our \KT\ \Ci\ planet candidates overlaid (red).
Also included is the location of the Sun (yellow) and host stars of previously
confirmed \KT\ planets (blue).
90\% of our candidates have photometry consistent with later spectral
types than the Sun.}
\label{fig:photometry}
\end{figure}

With the exception of one star in our sample (K2-18), we do not have
spectroscopic data with which to characterize the stellar properties.
Additionally, there are no measured parallaxes for any of these stars.
Instead, we rely on photometry.
For each system, we query the VizieR database of astronomical catalogs
\citep{Ochseinbein00}.
We record the $B$, $V$, $g'$, $r'$, and $i'$ magnitudes and their
uncertainties from the AAVSO Photometric All-sky Survey (APASS) DR6
\citep{Henden14}, as reported in the UCAC4 Catalog \citep{Zacharias12}.
We also record the $J$, $H$, and $K$ magnitudes and their uncertainties
as found in the 2MASS All-sky Catalog of Point Sources \citep{Cutri03}
and the $W1-W3$ \textit{WISE} magnitudes and uncertainties from the ALLWise Data
Release \citep{Cutri13}.
For all except two of our targets, the $W4$ band is only an upper limit,
and in the remaining two cases, the photometric uncertainity in $W4$ is at least an
order of magnitude larger than those in $W1-W3$, so we do not use $W4$
for any system.
These data are reported in Table 1, and a color-color diagram showing the
$r-J, J-K$ colors of our candidates is included as Fig.~\ref{fig:photometry}.

\subsection{Stellar Models}
\label{sec:stellarparams}
To convert the observed photometric data into physical properties for each
star, we used the new publicly available \texttt{isochrones} Python module,\footnote{
\url{http://github.com/timothydmorton/isochrones}} a general-purpose
interpolation tool for the fitting of stellar models to photometric or spectroscopic
parameters \citep{Morton15a}.
This software does trilinear interpolation in mass--age--\feh\ space for any
given set of model grids, thus being able to predict the value for any
physical or photometric property provided by the models at any values of
mass, age, and \feh\ within the boundaries of the grid.

This enables a set of observed properties ($\left\{ x_i, \sigma_i \right\}$),
either spectroscopic, photometric,
or both, to define a likelihood function to be sampled:
\begin{equation}
\label{eq:ischochronelhood}
\ln \mathcal L(\boldsymbol{\theta}) \propto -\frac{1}{2} \displaystyle \sum_i \frac{\left(x_i -
  I_i\left(\boldsymbol{\theta} \right)\right)^2}{\sigma_i^2},
\end{equation}
where $I_i(\boldsymbol{\theta})$ is the isochrone model prediction of property
$i$ at the given parameters $\boldsymbol{\theta}$.
If the observed properties include any apparent magnitudes, then
$\boldsymbol{\theta}$ includes distance and extinction in addition
to mass, age, and \feh.

In this work, we use grids from the Dartmouth Stellar
Evolution Database \citep{Dotter08} at Solar values of \afe=$0.0$ and
helium abundance $Y=0.2741$, which come packaged with the \texttt{isochrones}
module.
We then infer the stellar parameters using \texttt{MULTINEST} \citep{Feroz09},
an implementation of a multimodal nested sampling algorithm,
for each host star conditioned on the observed
photometric properties as presented in Table 1.
\texttt{MULTINEST} is designed to sample multimodal posteriors, where other 
samplers such as MCMC algorithms often struggle. 
Given the multimodal nature of our posteriors, this scheme is optimal for 
capturing parameter space on the subgiant branch where these stars could reside.
We include a prior on stellar metallicity representative of the observed metallicities
of stars within 1 kpc of the Sun, following the results of \citet{Hayden15}, and a Salpeter-slope
prior on mass up to the maximum mass available in the model grids of $3.7$ $M_\odot$.

During the sampling process, we fit for Galactic
extinction as one of our physical parameters.
We include the \textit{WISE} bandpasses by applying the relative extinction values between
SDSS, 2MASS, and \textit{WISE} calculated by \citet{Davenport14}.
In each step of our fitting process, we draw a value for $A_V$, calculate the expected extinction
in all bandpasses $A_X$ assuming the $R_V = 3.1$ reddening law of \citep{Fitzpatrick99}, 
and then measure the likelihood of our model stellar fit to the observed apparent magnitudes.
We apply a uniform prior ranging from zero to a maximum extinction value of 0.2 and
marginalize over extinction in our final determination of stellar parameters.
The NASA/IPAC Extragalactic Database, which reports the
\citet{Schlafly11} recalibration of the \citet{Schlegel98} extinction map as measured by
COBE/DIRBE and IRAS/ISSA, suggests that typical $A_V$ extinction values to the edge
of the galaxy at this
high Galactic latitude are $\sim$0.1 mag, so our upper limit appears to be justified.

Such a scheme enables us to infer the statistical uncertainties on the
mass, radius, and effective temperature.
However, we are subject to biases induced by systematics in the models themselves.
There is some evidence that the Dartmouth models may under-predict
radii of M dwarfs by $\sim 15\%$ when compared to other methods \citep{Newton15,
Montet15}.
Such an effect may be the result of the Dartmouth model reliance on BT-Settl
atmospheres, which are based on incomplete molecular line lists and
have been shown to predict near-IR colors that are too blue
\citep{Thompson14}.

As our stellar results are model-dependent, we caution users who intend to use these
parameters for other works, such as exoplanet population studies. When available,
stellar parameters inferred through other techniques such as asteroseismology or
spectroscopy should supersede these values.
We note the observed photometric parameters are consistent with spectroscopically derived
parameters for stars with published spectra, and consistent with typical
model-dependent uncertainties
from photometric data \citep[e.g.][]{Huber14}.
We provide full samples of our posteriors on the physical parameters for each
star.\footnote{\url{http://www.astro.princeton.edu/~tdm/k2/}}

\citet{Bastien14} use the ``granulation flicker'' in the \kep\ light curves
to suggest that
approximately $50\%$ of planet host stars have evolved off the
main sequence onto the subgiant branch, so that both the host stars and their
planets are larger than previously reported.
Similarly, in \KT\ \Ci\ we may expect to find evolved stars in a sample of planet
candidates, although we may expect the effect to be lessened due to the
high Galactic latitude of \Ci.
Indeed, we find this to be the case.
Two stars, EPIC 201257461 and 201649426 are definitively evolved stars,
with inferred masses less than 2 \msun but radii above 8 \rsun.
For approximately one third of the others,
we find the stellar radius posterior distribution to be bimodal, with both main
sequence and subgiant models of the stars being consistent with the photometric
data.
This number is consistent with our expectations of the number of subgiant
contaminants in the \Ci\ field (K. Stassun 2015, private communication).
Future observations to measure the parallaxes of these stars, such as with Gaia,
will be helpful in differentiating between these two models to determine
more precisely the stellar, and thus the planetary, radii.

\subsection{SuperNova Integral Field Spectrograph (SNIFS) and SpeX Spectroscopy}
\label{Spexobs}

A near-infrared spectrum of K2-18 was obtained using the upgraded SpeX
(uSpeX) spectrograph \citep{Rayner03} on the NASA Infrared Telescope Facility
(IRTF) on 2015 January 29 (UT).
SpeX observations were taken using the short cross-dispersed mode and the
0.3$\times15\arcsec$ slit, which provides simultaneous coverage from 0.7
to 2.5$\mu$m at $R\simeq2000$.
The target was observed at two positions along the slit to subsequently subtract
the sky background. Eight spectra were taken following this pattern, which provided
a final signal-to-noise ratio (S/N) of $>150$ per resolving element.
The spectrum was flat fielded, extracted, wavelength calibrated, and stacked
using the \textit{Spextool} package \citep{Cushing04}.
An A0V-type star was observed immediately after
the target, which was used to create a telluric correction using the
\textit{xtellcor} package \citep{Vacca03}.

An optical spectrum was obtained using SNIFS
\citep{Aldering02,Lantz04} on the University of Hawai'i
2.2m telescope on the night of 2015 January 30.
SNIFS provides simultaneous coverage from 3200--9700\AA\ at a resolution
of $\simeq1000$. Final S/N of the spectrum was $>100$ per resolving element
in the red ($\sim6000$\AA).
Details of the SNIFS reduction, including dark, bias, and flat-field corrections,
cleaning the data of bad pixels and cosmic rays, and extraction of the
one-dimensional spectrum are described in \citet{Bacon01} and
\citet{Aldering06}.
Flux calibration was performed using a separate pipeline described in \citet{Mann15}.

\teff\ was calculated by comparing our optical spectra with the CFIST
suite\footnote{\url{http://phoenix.ens-lyon.fr/Grids/BT-Settl/CIFIST2011/}} of the BT-SETTL
version of the PHOENIX atmosphere models \citep{Allard13}, which gave a temperature
of 3503 $\pm$ 60\,K.
More details of this procedure are given in \citet{Mann14} and
\citet{Gaidos14}.
This method was used because it is known to accurately reproduce empirical
\teff\ values from long-baseline optical interferometry \citet{Boyajian12}.

Metallcity was determined using the procedures from \citet{Mann13a}, in which the
authors provide empirical relations between atomic features and M dwarf
metallicity, calibrated using wide binaries.
We adopted the weighted mean of the $H-$ and $K-$band calibrations,
which yielded a metallicity of 0.09$\pm$0.09.

We combined the derived \teff\ and [Fe/H] values with the empirical
\teff-[Fe/H]-$R_*$ relation from \citet{Mann15} to compute a radius.
Accounting for measurement and calibration errors in [Fe/H] and \teff\ we calculated
a radius 0.394$\pm0.038R_\odot$.
We use these parameters instead of the derived photometric properties for this target,
although we note the two are consistent at the $1\sigma$ level.

The full list of stellar parameters adopted in this paper is included in
Table 2.

\section{Planet Properties}

In \paperit, only parameters directly observable from the \KT\ light curve
itself were reported: the period, time of transit center, and transit depth.
With stellar properties now in hand, we can convert these observational
results into fundamental parameters of each planet candidate.
For each candidate, we fit the light curve using a physical transit model
\citep{Mandel02, Kipping10b} simultaneously with a systematics model similar
to the one described by \paperit.
We use \texttt{emcee} \citep{Foreman-Mackey12}, an implementation of the
affine-invariant ensemble sampler of \citet{Goodman10}
 to sample from the posterior probability distribution
for the stellar---limb darkening coefficients, mass, radius, and effective
temperature---and planetary---radius, period, phase, impact parameter,
eccentricity, and argument of periapsis---parameters, conditioned on the
light curve and the measured stellar properties.

Following \paperit, the likelihood function that we use is marginalized over
the weights of the ``eigen light curves'' in the linear systematics model.
Unlike \paperit, we include an empirical Gaussian prior on the weights
determined by robustly computing the distribution of weights across the full
set of Campaign~1 light curves.
This prior mitigates the incorrect detection of false signals induced by
stellar variability---as discussed below in Section~\ref{sec:systematics}---so
we exclude these candidates (EPIC 201929294 and EPIC 201555883) from the tables
of results.

In this analysis, we assume the dilution caused by additional stars
contributing flux into the aperture is negligible for nearly all systems.
Given the location of the \Ci\ field at a high Galactic latitude, we expect
low contamination by background giants.
Nevertheless, this assumption may not be valid for all systems.
Any contamination unaccounted for, as may happen if any of these stars are
actually unresolved binaries, would cause us to underestimate the radii of any
planets we detect.
Therefore, high-contrast adaptive optics imaging of any systems should be
obtained before these planets are used in population inference studies.
The planet parameters measured by this analysis are listed in Table 3.

\section{False Positive Analysis}

There are many scenarios which can cause an astrophysical false positive,
where an EB star masquerades as a transiting planet.
The most common scenarios are if (a) it is a highly grazing eclipse, or (b) the binary system
shares a photometric aperture with a significantly brighter star,
resulting in a diluted eclipse depth.
When possible, such astrophyscial false positive scenarios are traditionally
ruled out by detailed follow-up observations, often a combination of
high-resolution imaging and radial-velocity measurements.
However, the \kep\ mission, with its thousands of planet candidates
around mostly faint stars, necessitated a paradigm shift---a move
toward probabilistic interpretation of transit signals, rather than
comprehensive follow-up of each individual candidate \citep{Morton11b}.

\citet{Morton12} presented an automated method to calculate the
probability that a planet candidate might be caused by an
astrophysical false positive.
This method uses Galactic population simulations to determine the
distributions of possible false positive scenarios, comparing the
typical light curve shape of each to the data.
It then combines this information with observationally motivated prior
assumptions about the populations of field stars, the properties of
multiple star systems, and the occurrence rate of planets as determined
from \kep\ \citep{Fressin13}, in order to determine the probability that
the observed signal may be a false positive.
Similar in spirit to other published methods of
probabilistic validation, such as BLENDER \citep{Torres11a} and PASTIS
\citep{Diaz14a}, it has the advantage of being computationally less demanding
and fully automated, and thus easily applied in batch to a large
number of candidates.

In this work, we use
\texttt{vespa}\footnote{\url{http://github.com/timothydmorton/vespa}} \citep{Morton15b},
a new publicly available, general-purpose implementation
of the \citet{Morton12} procedure,
to calculate FPPs for each of these
\KT\ candidates.
The following constraints on false positive
scenarios are imposed:
\begin{itemize}
\item A chance-aligned EB system may reside anywhere inside
or within one pixel of the photometric aperture of the target star.
In creating a light curve for each star, we define photometric apertures ranging
from 10 to 20 arcsec for each star, as defined in Table 4.
Given the 6 arcsec point spread function (PSF) of the \kep\ telescope, we allow for the possibility
that companions falling just outside of our aperture (within one pixel) may
contribute to the light curve, possibly causing a false positive event.
The search for such companions is discussed in
\textsection \ref{sec:background}.
\item The maximum allowed depth of a potential secondary eclipse event
is the most significantly detected signal at the same
period of the planet candidate, once the primary transit is masked out
(discussed in \textsection\ref{sec:sec}).
\texttt{vespa} does not allow for the possibility of secondary eclipses
larger than those observed in the \KT\ light curve for each star.
\item Blended stars must be allowed by the available adaptive optics
and archival imaging data (discussed in detail in \textsection
\ref{sec:AO} and \ref{sec:background}).
\texttt{vespa} only considers stars below the detection threshold for
the AO imaging, which is a position-dependent value following a calculated
contrast curve for each star.
\end{itemize}

Each of these scenarios is an astrophysical eclipse, caused by one
object passing in front of another, blocking some fraction of the
total light.
The calculations here do not include the possibility
that each signal is caused by an instrumental artifact in the data or
some other astrophysical event, such as stellar activity, masquerading
as planet transits.

Table 5 summarizes the results of these
calculations, presenting the relative probability for each candidate
to be caused by any of three false positive scenarios: an undiluted
EB, a hierarchical triple eclipsing binary (HEB),
and a chance-aligned background(/foreground) eclipsing binary (BEB).

Six of the presented candidates have FPP $>$90\%;
these are considered to be likely false positives.
On the other hand, 24 candidates
have FPP $<$ 1\%.
Three of the transit signals might plausibly be caused by
contamination by detected stellar companions within the photometric apertures
(see \S\ref{sec:background}), so we keep these as candidates.

This leaves 21 candidates that we statistically validate as planets, including
four that have been previously
identified in the literature \citep{Armstrong15b, Crossfield15}.
So in total, of the 36 candidates, 21 are secure planets,
17 of which we validate here for the first time.

\begin{figure*}[htbp]
\centerline{\includegraphics[width=1.0\textwidth]{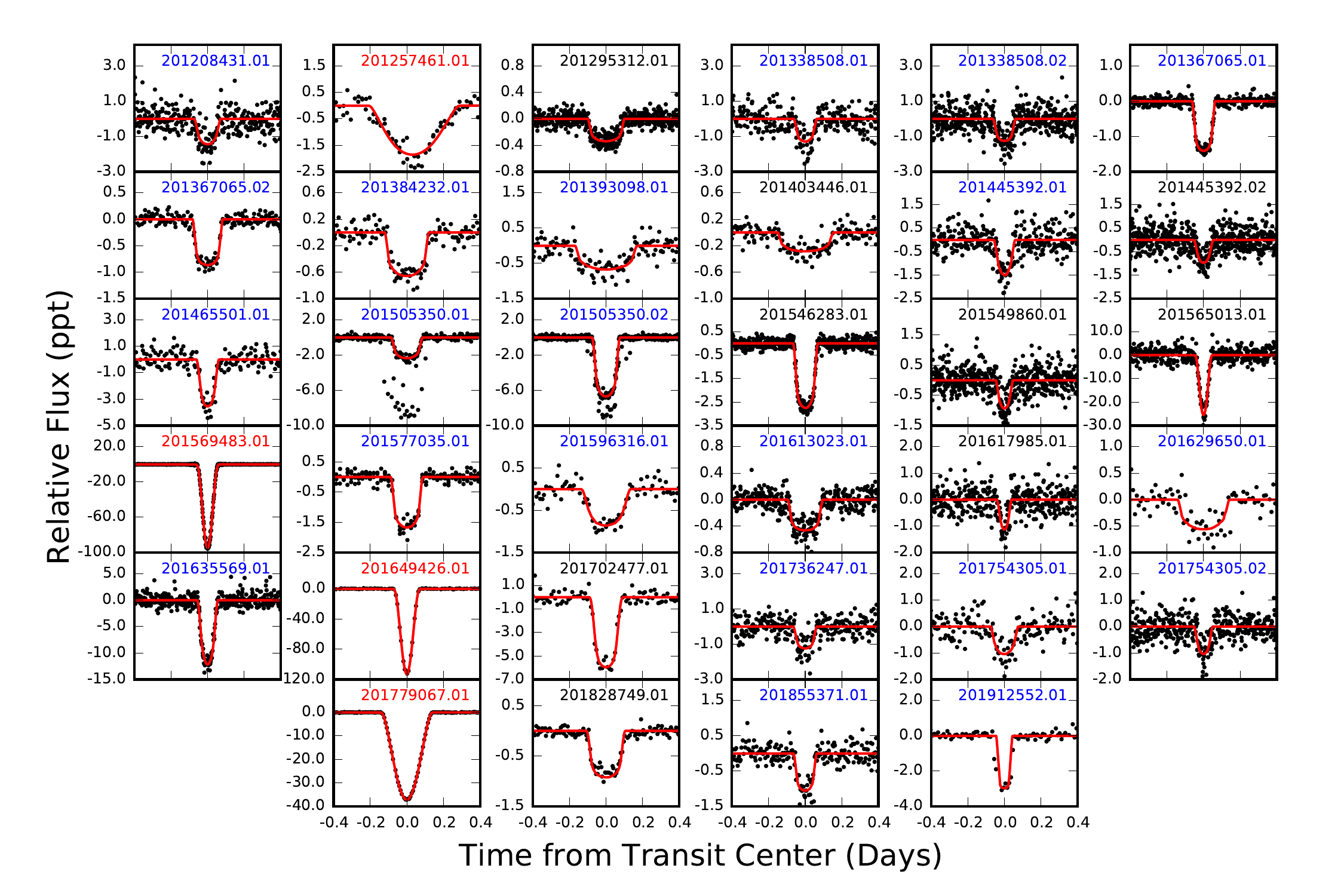}}
\caption{Phase-folded \KT\ photometry for all planet candidates analyzed in
this paper.
Each is the product of a fiducial noise model, in which the median systematic
has been removed for illustrative purposes.
The systems which we validate as transiting planets are labeled in
blue.
The systems which we confirm as false positive events are labeled in red.
The systems which we leave as candidates are labeled in black.
Red curves outline the median transit model for each candidate system.
}
\label{fig:transits}
\end{figure*}

We emphasize that the majority of these validations rely
solely on the transit photometry and SDSS data, with follow-up imaging only obtained for
seven of the 31 targets.  This
demonstrates the utility of the \texttt{vespa} tool, which will be
crucial to interpreting future candidates detected by \KT, \textit{TESS}, and \textit{PLATO}
and prioritizing follow-up observing efforts.
We show the transit signals in Figure~\ref{fig:transits}.

\subsection{Secondary Eclipse Observations}
\label{sec:sec}
One of the definitive signatures of a false positive binary star system masquerading
as a transiting planet is the presence of a secondary eclipse. While a nondetection of a secondary
does not exclude the possibility of a binary system (the orbit may be eccentric,
or the companion too faint for a secondary eclipse to be detectable in the noise),
such a nondectection reduces the probability of each of the EB false
positive scenarios.

To attempt to eliminate each EB scenario, we first search each \KT\
light curve to determine which secondary eclipse signals are not allowed by the data.
We mask the transit signal of the planet in question and search for the most significant
signal at the same period.
Such a scheme does not assume circular orbits: we return the most significant signal
at any phase, not only at the midpoint between consecutive transits.

We report these maximum allowable secondary eclipse depths in Table 5.
These values are used by \texttt{vespa} as limits on the allowable secondary eclipse.
Any models that cause a larger event, such as a background EB
consisting of two equal-mass stars in a circular orbit, can be excluded by the data.
We note that with the exception of K2-19c, all systems with a maximum
eclipse depth of at least one part per thousand have FPPs of 0.866 or larger.
The exception, K2-19, is a two-planet system with the two planets near a 3:2 period
commensurability, so in this case the ``secondary'' is actually the transits of the
other planet.

\subsection{Adaptive Optics Imaging}
\label{sec:AO}

We obtained high resolution images of seven stars with the
Palomar High Angular Resolution Observer (PHARO) infrared detector
\citep{Hayward01} behind the PALM3000 adaptive optics system \citep{Dekany13}
at the Palomar 5.1 m Hale telescope on the nights of 2015 February 3 and 4 UT.
Sky conditions were mostly clear with light cirrus and $\approx$1$\farcs$0--1$\farcs$3 seeing
on both nights.
We used the smallest plate scale of 25~mas pixel$^{-1}$ which resulted
in a field of view of 25$\farcs$6$\times$25$\farcs$6 across the 1024$^{2}$ pixel$^{2}$ array.
All observations were obtained with the 32x pupil sampling mode, resulting in Strehl ratios
of $\approx$20--30\% in $K_S$ for our $V$=11--13~mag targets as measured by the Strehl
monitor at the telescope in real time.
We obtained unsaturated dithered frames of each target in $K_S$-band with typical
integration times of 2--10~s.
Except for EPIC~201828749 and EPIC~201546283, which had nearby candidate
binary companions, we also
acquired deep saturated images (5--10 frames at 60 s each)
to search for fainter companions.

Images were registered and contrast curves were generated following \citet{Bowler15a}.
For the saturated data, the star's position in each image was found by masking the
saturated region and fitting a 2D bivariate Gaussian to the PSF wings.
Contrast
curves for the median-combined image are calibrated using the unsaturated frames.
The typical sensitivity is 6.5--7.5~mag at 1$''$.
The images were astrometrically calibrated using dithered observations of the
Trapezium cluster centered
on $\theta$$^1$~Ori~C taken on 2015 February 3 UT.
Based on the reference astrometry for pairs of stars in the field from
\citet{McCaughrean94}, we measure a plate scale of 25.2~$\pm$~0.4~mas pixel$^{-1}$
and north orientation of --$0\fdg2~\pm~0\fdg3$.
Since this latter value is consistent with being aligned with
the detector columns, we adopt a value of $0\fdg0~\pm~0\fdg3$ for this work.
Relative photometry of nearby stars is carried out using aperture photometry with
an aperture radius of 12~pixels (0$\farcs$3).
For EPIC~201828749, we also acquired $J$- and $H$-band images.
Astrometry and photometry is derived separately for each image, and the mean and standard deviation
of these measurements is adopted for our final values listed in Table~4.

\begin{figure*}[htbp]
\centerline{\includegraphics[width=1.0\textwidth]{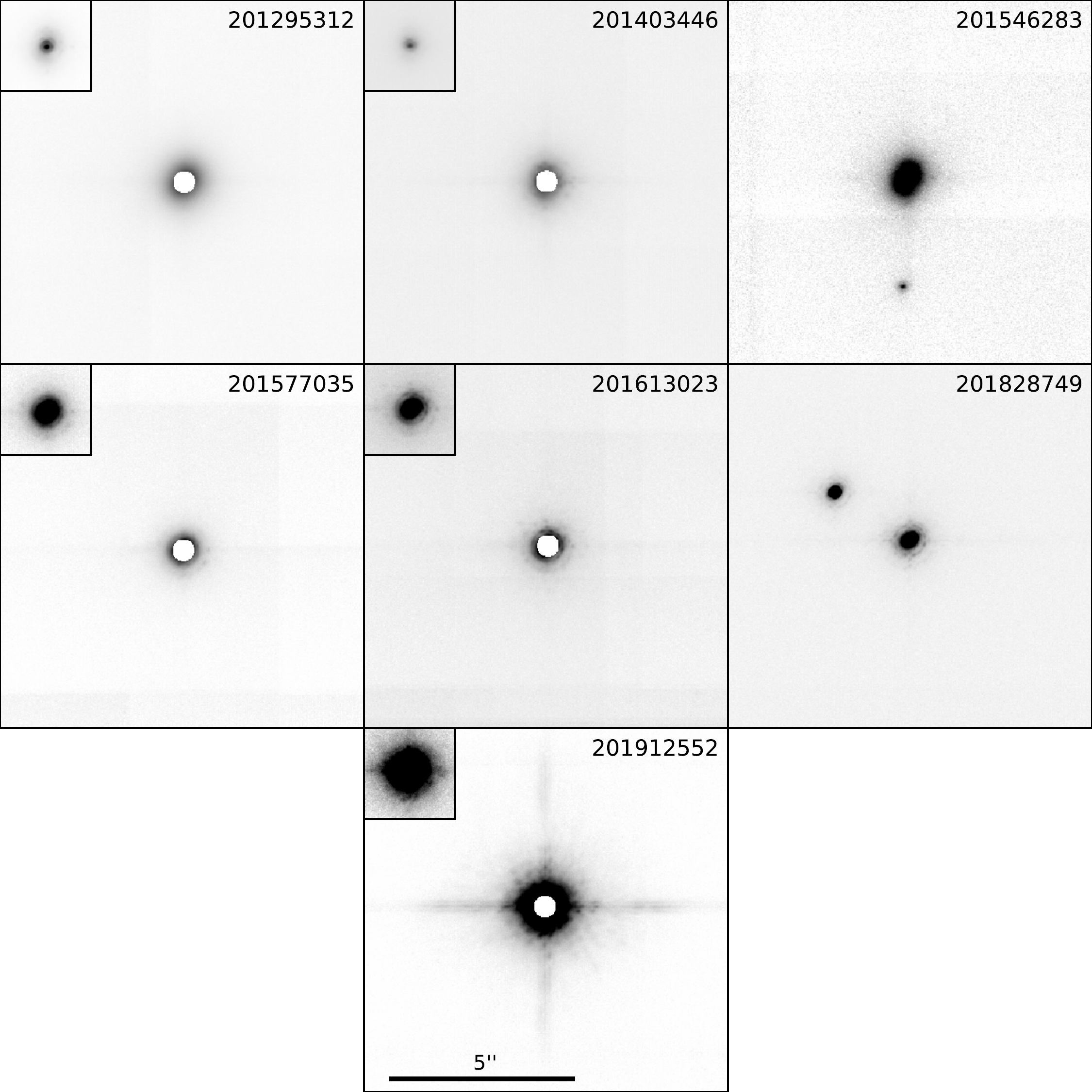}}
\caption{Adaptive optics images for the seven stars observed with high-contrast imaging.
The main frame for each single system shows the deep, saturated image.
The inset for each single system shows a shallower, unsaturated image to better
identify companions at close projected orbital separations.
For the two systems with imaged companions, EPIC 201546283 and
EPIC 201828749,
only unsaturated frames are collected.
The pixel scale is $0\farcs0252$ per pixel.
Each subplot is a square 400 pixels on a side and
each inset is a square 100 pixels on a side.
All subplots, including insets, are plotted on the same scale.}
\label{Fig:AO}
\end{figure*}

\begin{figure}[htbp]
\centerline{\includegraphics[width=0.5\textwidth]{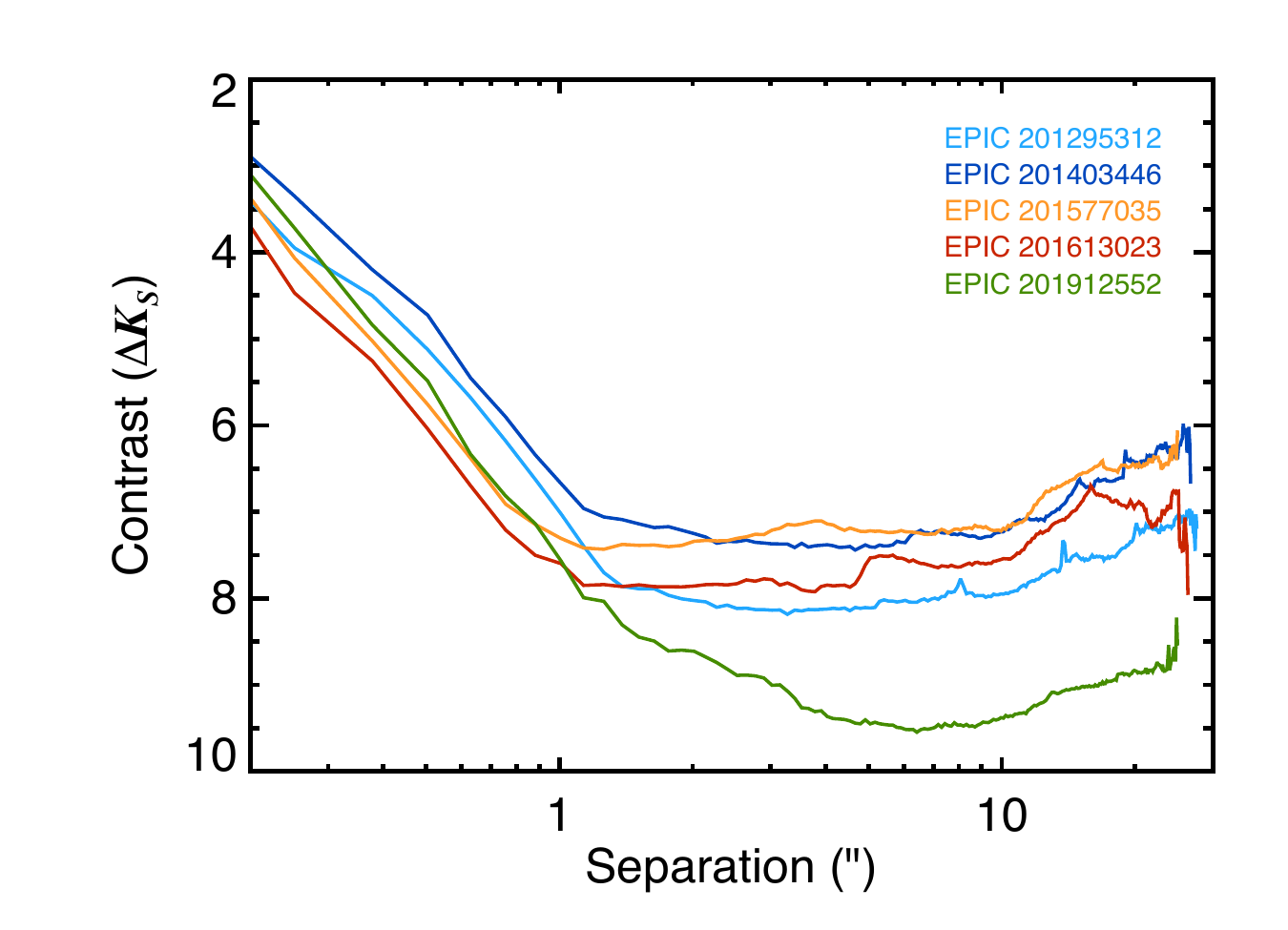}}
\caption{$5\sigma$ contrast curves for all systems with AO nondetections.
For all systems, we can exclude the possibility that a companion at a given $\Delta K_S$
exists.
From our known transit depths, we can then rule out significant parameter space in which
an eclipsing binary could reside and mimic a transit signal.}
\label{Fig:CC}
\end{figure}

Images for all systems AO data was obtained for is shown in Figure~\ref{Fig:AO},
while contrast curves showing the $5\sigma$ limits for detection
as a function of orbital separation are given in Figure~\ref{Fig:CC}.

\subsection{Known Background Stars}
\label{sec:background}
The PHARO AO system has a field of view of 25 arcsec.
Each \KT\ pixel is a square, $3\farcs98$ on a side.
A background EB within a few \KT\ pixels of our target
stars could mimic a transit signal inside our aperture while evading detection
by PHARO.
Such wide EBs should appear in seeing-limited ground-based
surveys.

To investigate the possibility that such wide companions exist,
we query the ninth data release of the Sloan Digital Sky Survey
\citep[SDSS DR9,][]{Ahn12}.
For each target, from the depth of the observed transit we determine how
bright a background object must be to cause the event if the background
object were an equal mass totally eclipsing binary.
We then search for all stars within 25\arcsec\ that are within this brightness
limit relative to the candidate host star.
All apertures we use in our \KT\ analysis are smaller than 20\arcsec\ so this
search should encompass the region where possible background
contaminants could reside.
Of the 31 stars in our sample, eleven have such a companion, plus one
detected in AO imaging.

Unlike the original \kep\ field, the field for \KT\ \Ci\ is
well out of the Galactic plane, so the rate of giant, distant background
stars is significantly lower.
We include all potential contaminants in Table 4.
We validate or eliminate each of these as a possibility based on the transit shape.
For example, the events near EPIC 201546283 could only be caused by a
background binary if the background object was a completely eclipsing system
(so that the eclipse depth was $50\%$).
In this case, the transit would be V-shaped.
Since it is not, the background object likely does not cause
the transit event.

In Table 4, the ``maximum depth'' column represents
the maximum observed ``transit'' depth if the transit were actually caused
by a total eclipse of the hypothetical background binary system, inducing
a 50\% flux decrement in the background star's apparent brightness.

The photometric apertures used to detect these candidates range in radius from
$10\farcs0$ to $19\farcs9$.  In order to be a plausible contaminant,
any companion star must
be either within this aperture or just outside but bright
enough for signficant flux to leak in.
Evaluating each of the systems listed in Table 4, we judge that we cannot yet
rule out contamination as a potential source of the transit signal for four
candidates: 201295312.01, 201403446.01, 201546283.01, and 201828749.01.
Despite receiving
low FPP scores from \texttt{vespa}, we list these systems as candidates in
Table 5, rather than planets.  Further updates to the \texttt{vespa} code
will allow consideration of ``specific'' false positive scenarios; that is,
scenarios that correspond to actually detected stars such as these, rather than
hypothetical background or bound companions.

The candidates with identified companions that we judge to not be plausible
sources of potential contamination are the following:
\begin{itemize}

\item K2-13b (201629650.01)--- The companion to this star is $17\farcs3$ from
the EPIC target.  As this is outside the aperture
(radius $15\farcs9$) and the background star is not particularly bright,
we rule out contamination for this system.

\item 201702477.01--- The companion to this star is $12\farcs15$ from the
EPIC target, and the aperture size is $10\farcs0$.  In addition,
the maximum depth in this system is almost identical to the transit depth.
For these two reasons we rule out contamination in this case.

\end{itemize}

SDSS is 95\% complete at $r=22.2$ mag and the telescope has a PSF of $1\farcs4$.
For the purposes of the \texttt{vespa} calculation, we thus
treat nondetection in SDSS data as providing a contrast curve at
wide separations
down to a limiting magnitude of $r=22.2$ mag.

\subsection{Archival Imaging}

For the stars with AO nondetections, there is still the
possibility that a background binary could be positioned
directly behind the target star, evading detection.
The probability is small, given the $0\farcs1$ diffraction limit
of the Hale Telescope at $2$ $\mu$m, but nonzero.
While the \texttt{vespa} calculations quantify this probability for
this to occur, we can also rule out the possibility of such
chance alignments, down to a certain contrast, with archival imaging data.

\begin{figure*}[htbp]
\centerline{\includegraphics[width=0.65\textwidth]{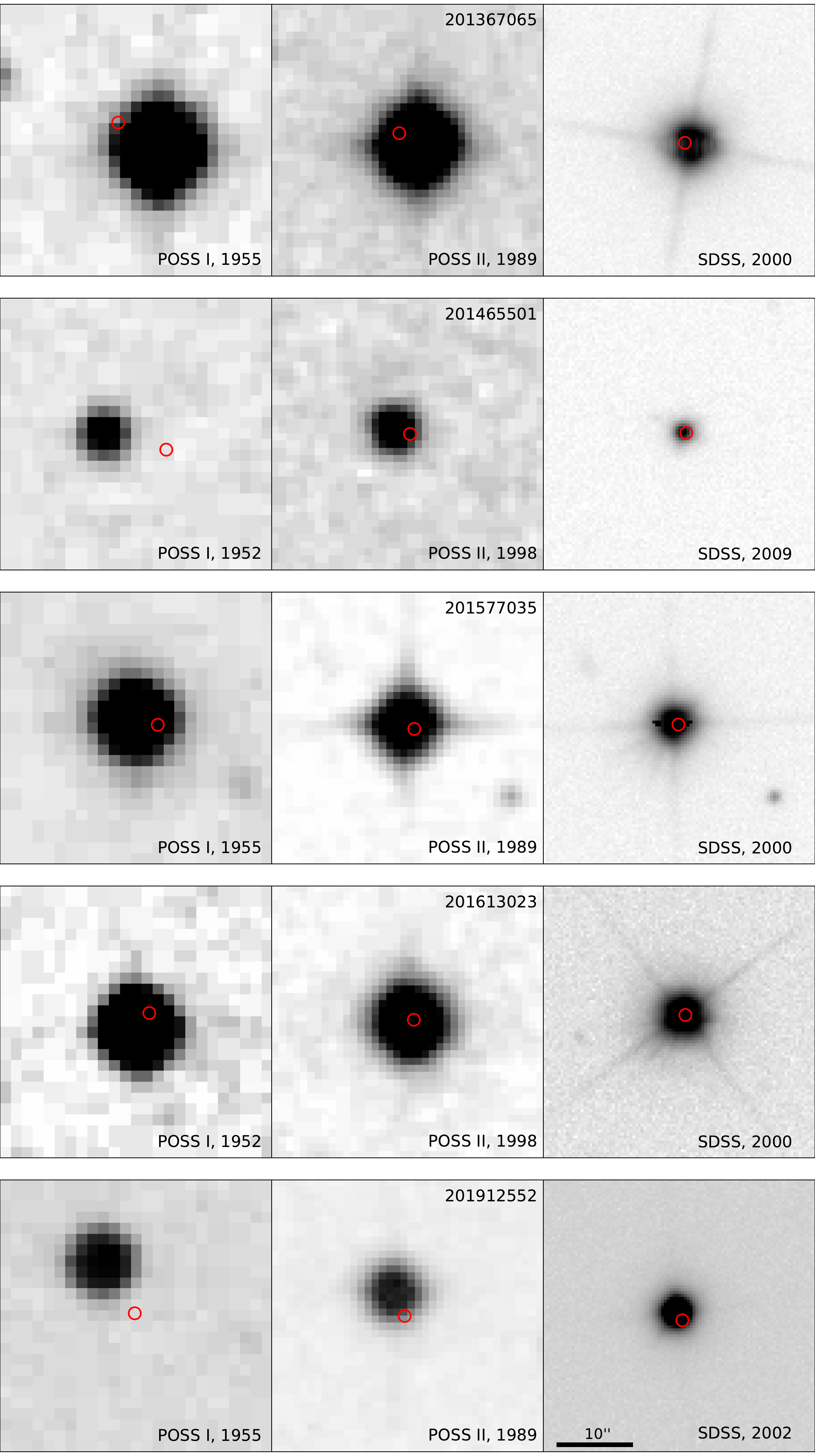}}
\caption{Archival maging for the five highest proper motion targets in our sample.
In all cases, there are no background
objects directly behind the present day location of the target (red circle) that could be missed
by the AO observations. Modern SDSS imaging can also rule out wide companions
that may have been missed at wide separations, beyond the AO field of view, such as the
companion which can be seen in the images of K2-10.
All figures are aligned such that north is up and east to the left.
All subplots are on the same scale.}
\label{fig:archival}
\end{figure*}

Five of the stars in our sample have proper motions larger than 50 mas
yr${-1}$, so they have moved across the sky by $\gtrsim 2\farcs5$ since
they were imaged during the first Palomar Observatory Sky Survey (POSS) in the
1950s.
To rule out background companions, we download data from the POSS I
and II surveys, which imaged these targets in 1952-1955 and 1989-1998,
respectively.
We also download data from the Sloan Digital Sky Survey, which imaged
these fields between 2000 and 2009.
As shown in Figure~\ref{fig:archival}, we do not detect any background
targets at the present-day location of any of these stars in any of these
images.

For these targets, we can extend our contrast curves to zero
present-day orbital separation and rule out the possibility that
these transit events are caused by a background EB.
By combining present-day seeing-limited photometric survey data,
adaptive optics imaging, and archival photometry, the only stellar companions
we would not detect would be those that are gravitationally bound to the target
star and positioned in their orbits so that their projected separation is
smaller than the diffraction limit of the Hale Telescope.
Such an alignment would require the orbital inclination of the binary to be
nearly $90^\circ$ and the phase $\varpi + \theta \approx \pi/2$ or $3\pi/2$.
While we cannot fully rule out this possibility, the \texttt{vespa}
calculations confirm that its probability is negligibly small.

\subsection{Tillinghast Reflector Echelle
Spectrograph (TRES) Radial Velocities}

We observed K2-18 on 2015 February 04 and 25 UT
with TRES on the 1.5 m Tillinghast Reflector at the Fred L.
Whipple Observatory.
These dates were chosen to be near the times of largest RV variations,
corresponding to phases of 0.72 and 0.32 relative to the time
of transit.
The spectra were taken with a resolving power of $R=44,000$ and
integration times ranging from 2800 to 3600 s, resulting in
S/N between 17 and 29 per resolution element.

The spectra were extracted as described in \citet{Buchhave10}.
The relative RVs were derived by cross-correlating the spectra against the
strongest observed spectrum (in this case, the first) over the wavelength
range 4700 - 6800 \AA.
We selected 19 echelle orders in the analysis, being careful to reject
orders with telluric absorption lines, fringing in the far red and those
with very low SNR in the blue.

The two observed spectra have RVs that differ by $47 \pm 42$ m s$^{-1}$.
If the RVs were caused by a stellar companion, the RV shift between these
observations would be on the order of km s$^{-1}$.
Therefore, we can rule out any stellar-mass companions that would be able
to create this transit signal.

\section{Potentially Interesting Systems}
\subsection{A Mini-Neptune with Earthlike Insolation}

The planet orbiting K2-18 may be an interesting target for
atmospheric studies of transiting exoplanets.

By combining archival and modern seeing-limited data with adaptive optics
imaging, we can exclude the possibility these transit events are caused by
a background EB.
The apparent transits must be caused by an object co-moving with K2-18;
radial velocities eliminate the possibility the companion is nonplanetary.
Therefore, we confirm the planetary nature of this system.

This star is an M2.8 dwarf at a distance of $34\pm4$ pc.
Of our planet candidate hosts, only K2-3 \citep[originally
discovered by][]{Crossfield15} is brighter in $K$-band.
This star is only 0.1 mag fainter in $K$ than GJ\,1214
\citep{Charbonneau09}.
Due to the relative brightness of the host star, this target is likely
to become a prime target for atmospheric characterization studies
and is ideal as a target for future space-based missions such as
\textit{JWST}.

The planet is slightly smaller than GJ\,1214b, but unlike that planet,
K2-18b is not highly irradiated.
Instead, it is at a reduced semimajor axis \ars$ = 83.8 \pm 9.0$.
Its equilibrium temperature is then, assuming zero albedo, $T_{eq} = 272 \pm 15$
K, meaning its bulk insolation is $94 \pm 21$ percent that of the Earth's.
Although the planet is likely too large to be rocky \citep{Rogers14},
its atmosphere is likely to be the focus of many future observations, providing
a cool analogue to the highly irradiated planets of a similar size found by
\kep.

\subsection{Other Sources of False Positives}
\label{sec:systematics}

The method of \paperit\ assumes that all variability in the light curves are
caused by either the motion of \KT, in which case the variability is shared
by all stars, or transits of planets, in which case the variability is intrinsic
to only one star.
This assumption breaks down for extremely spotted stars where the
astrophysical variability is larger than the instrumental
magnitude.
In that regime, the starspot modulations can be incorrectly fit by the
systematic model, causing spurious transits to appear.
This appears to be the case with EPIC 201929294, which has
coherent starspots that appear to have the same rotation period as the
transit period reported previously.
Because the starspots are so periodic and coherent, these spurious
transits were falsely identified as a planet candidate; we consider that
system a false positive in this work.

The candidate object possibly orbiting EPIC 201555883 has a period,
time of transit, and transit duration consistent with EPIC 201569483.
Such effects are not uncommon in \kep\ data.
\citet{Coughlin14} identify 685 KOIs as false
positives and outline four physical reasons why these anomalies may occur.
EPIC 201555883 is a unique case in that it does not appear to fall under any
of these cases. It falls on module 23, while EPIC 201569483 is on module 8,
neither $180^{\circ}$ away from nor on the same column as this candidate.
Moreover, there is not any evidence of a mechanism that could cause a third
star to induce both the appearance of a $7\%$ eclipse on one module and an
additional anomalous transit event on a different module.
Instead, this candidate could be a false positive caused by a different
systematic mechanism.

\paperit\ modeled the systematic effects in the \KT\ light curves using a
linear combination of ``eigen light curves'' generated empirically by running
a principle component analysis on the light curves of every star.
This means that the training set includes the light curves for variable stars,
EBs, and even transiting planets.
Again, this star has significant variability caused by starspots.
In this case, the fitting procedure tries to account for stellar variability using
the eigen light curves.
This overfit gives undue weight to eigen light curves that include the transits
of EPIC 201569483, causing this spurious transit to occur.
Again, we consider this system to be a false positive.
As stated in Section 3, by including an empirical Gaussian
prior on the weights for the eigen light curves in the linear systematics model,
the signals observed on EPIC 201555883 and 201929294 are mitigated, suggesting
such a scheme should be employed in searching for planet candidates in future
campaigns.

The problem of over-fitting stellar variability using eigen light curves can also be
solved by adding a stellar activity model to our fitting procedure.
In this case, the spacecraft motion could be fit simultaneously with a model
of starspot modulation, astroseismic oscillations, and planet transits.
Such a model is currently under development (Angus et al. 2015, in preparation)

\subsection{Multiple Planet Systems}
Five of the systems reported by \paperit\ have more than one transiting
candidate.
One of these is K2-3, a three-planet system originally announced by
\citet{Crossfield15}.
Another of these is K2-19 \citep{Armstrong15b}, a two-planet system with the
orbital periods of the two planets near a 3:2 period commensurability.
The remaining three are all representative of the multiple-planet systems observed by
\kep\ \citep{Lissauer11b, Fabrycky14}.
Two of the systems are near a period commensurability and all three consist of
mini-Neptune sized planets.

We do not detect any significant transit timing variations (TTVs) in any of these systems from
the \KT\ data alone.
K2-5 would be expected to have a TTV period of 117 days, but
is likely too far from commensurability to have an observable TTV signal.
K2-8 is expected to have a TTV period of 234 days, so this system may be a candidate
for additional follow-up to constrain the system masses dynamically.
The transiting planets orbiting K2-16 are near a 5:2 period commensurability.
There is no evidence from \kep\ of an abundance of planets near this period ratio, and so this
may be coincidence.
Follow-up observations may be warranted to search for an additional planet in this system
forming a resonant chain, similar to those observed around other stars \citep[e.g.][]{Swift13,
Campante15}.

\subsection{Systems Orbiting Bright Stars}

One of the primary goals of \KT\ is the detection of transiting planets around bright stars that
can be followed up from the ground or with future space-based observatories such as JWST
\citep{Howell14}.
Of our sample, two systems orbit stars with $K < 9$ mag: K2-3 \citep{Crossfield15} and
K2-18.
An additional planet candidate may orbit EPIC 201828749, a star with $K = 9.93 \pm 0.03$ mag.
These targets are ideal for ground-based followup and may be useful targets for \textit{Spitzer} and
\textit{JWST} to probe planetary atmospheres.

\section{Results and Discussion}

We have presented stellar parameters for all planet candidates systems identified by
\paperit.
We statistically validate \Nvalidated\ of the 36 candidates as bona fide planets, and we
identify \Nfp\ as false positives, including two systematic false alarms.
Of the planets, 4 have been previously validated in other works, while \Nvalnew\ are
validated here for the first time.
The systems not validated as planets or false positives remain as planet candidates.

Enabling much of this analysis are two new open-source Python packages:
\texttt{isochrones}\footnote{\url{http://github.com/timothydmorton/isochrones}},
which we use to infer posteriors on physical stellar properties based
on fitting theoretical stellar models to observed data; and \texttt{vespa}\footnote{
\url{http://github.com/timothydmorton/vespa}}, a
new implementation of the \citet{Morton12} transit false positive analysis scheme.
Both of these packages will continue to be useful in future analysis of transit candidates
where comprehensive follow-up observations may be unavailable.

The \texttt{isochrones} package uses the nested sampling scheme 
\texttt{MULTINEST} to capture the true multimodal nature of the posteriors.
Using an MCMC algorithm instead can cause only one peak in the posterior
distribution to be sampled. 
If the photometry is consistent with both a star on the main sequence and the
subgiant branch, an MCMC technique could cause one of these peaks (likely the
subgiant possibility) to be missed, leading to an underestimation in the likelihood
of subgiant stars and and underestimation of the uncertainties of both the stellar and planetary 
parameters.

With the exception of one object, all of the stellar parameters are derived from comparing
photometric observations to the Dartmouth stellar evolution models.
As a result, both the stellar and planet parameters are subject to systematic biases induced
by discrepancies between the models and reality.

The planets we confirm in this paper, like the planets found in the original \kep\ mission,
span a wide range of parameter space.
They are at distances ranging from $34$ to $700$ pc, have radii ranging from $1.3$ to
5.3 $R_\oplus$, and orbit with periods ranging from 5.0 to 50.3 days.
Like the original mission, we find significantly more small planets than large planets, as
expected from the radius distributions measured from \kep\
\citep{Howard12, Fressin13, Morton14}.

Unlike the original mission, however, we find that nearly all of our confirmed planets
are around stars less massive than the Sun.
This difference is a result of both the \Ci\ field and the target selection process.
\Ci\ is at a significantly higher Galactic latitude than the original \kep\ mission,
meaning there is a much lower number density of targets at large distances.
As massive stars
at kiloparsec distances are relatively less likely to exist in \Ci\ than near the
Galactic plane, the pool of targets that could be selected for \Ci\ contains a larger
fraction of subsolar stars.

Low-mass stars, particularly M dwarfs, are also a specific focus of the \KT\ mission.
One of the primary goals of the \kep\ mission was to ``determine the abundance of
terrestrial and larger planets in or near the habitable zone of a wide variety of spectral
types of stars'' \citep{Batalha13} However, $\sim70\%$ of Kepler's target stars had
masses within 20\%
of the Sun's, while 70\% of the stars in the Galaxy have less than 50\% the mass of the
Sun \citep{Brown11}.
\KT\ will fulfill the promise of \kep, with the goal of providing a yield of small planets
around bright, small stars to facilitate follow-up measurements \citep{Howell14}.
This is clear from the \KT\ target selection process, with thousands of K and M dwarfs
being selected in each campaign.
Based on these plans, we expect that \KT\ will detect hundreds of planets during its
lifetime, with the majority being mini-Neptunes and super-Earths around stars
less massive than the Sun.

\acknowledgements
We thank Roberto Sanchis-Ojeda (Berkeley), Dan Huber (Sydney) and
Jeff Coughlin (SETI) for
conversations and suggestions which improved the quality of this manuscript.
We also thank Keivan Stassun (Vanderbilt) for his insights into
 stellar parameters and the rate of subgiant contamination for both \kep\ and \KT, 
 which significantly improved this work.
We thank the anonymous referee for their comments and suggestions.

We are grateful to the entire Kepler team, past and present.
Their tireless efforts were all essential to the tremendous success of the mission and the
successes of \KT\, present and future.

Some of the data presented in this paper were obtained from the Mikulski
Archive for Space Telescopes (MAST).
STScI is operated by the Association of Universities for Research
in Astronomy, Inc., under NASA contract NAS5--26555.
for MAST for non--\textit{HST} data is provided by the NASA Office of Space
Science via grant NNX13AC07G and by other grants and contracts.

This paper includes data collected by the \kep\ mission.
Funding for the \kep\ mission is provided by the NASA Science
Mission directorate.

This paper includes data collected by the Sloan Digital Sky Survey.
Funding for SDSS-III has been provided by the Alfred P. Sloan Foundation,
the Participating Institutions, the National Science Foundation, and the
U.S. Department of Energy Office of Science. The SDSS-III web site is
http://www.sdss3.org/.
SDSS-III is managed by the Astrophysical Research Consortium for the
Participating Institutions of the SDSS-III Collaboration including the
University of Arizona, the Brazilian Participation Group, Brookhaven
National Laboratory, Carnegie Mellon University, University of Florida,
the French Participation Group, the German Participation Group, Harvard
University, the Instituto de Astrofisica de Canarias, the Michigan
State/Notre Dame/JINA Participation Group, Johns Hopkins University,
Lawrence Berkeley National Laboratory, Max Planck Institute for
Astrophysics, Max Planck Institute for Extraterrestrial Physics, New
Mexico State University, New York University, Ohio State University,
Pennsylvania State University, University of Portsmouth, Princeton
University, the Spanish Participation Group, University of Tokyo,
University of Utah, Vanderbilt University, University of Virginia,
University of Washington, and Yale University.

B.T.M. is supported by the National Science Foundation Graduate Research
Fellowship under Grant No. DGE-€1144469.
J.A.J. is supported by generous grants from the David and Lucile Packard
Foundation and the Alfred P. Sloan Foundation.

D.F.M. and D.W.H. were partially supported by the National Science Foundation
(grant IIS-1124794), the National Aeronautics and Space Administration (grant
NNX12AI50G), and the Moore-Sloan Data Science Environment at NYU.

T.D.M.~is supported by the National Aeronautics and Space Administration (grant
NNX14AE11G).

{\it Facilities:} \facility{Kepler},  \facility{PO:Hale (PHARO)}, \facility{FLWO:1.5m},
\facility{IRTF:SpeX}

\bibliography{exopapers}

\include{table_photometry}
\include{table_stellarprops}

\begin{turnpage}
\include{table_planets}
\end{turnpage}

\include{table_bg}

\include{table_fpp}

\end{document}

%% file: table_photometry.tex
\clearpage
\begin{turnpage}
\begin{deluxetable*}{cccccccccccc}
\tablewidth{0pt}
\tabletypesize{\scriptsize}
\tablecaption{Photometry for all Objects of Interest \label{Tab:Photometry}}
\tablehead{
\colhead{EPIC} &
\colhead{$B$\tablenotemark{1}} &
\colhead{$V$\tablenotemark{1}} &
\colhead{$g$\tablenotemark{1}} &
\colhead{$r$\tablenotemark{1}} &
\colhead{$i$\tablenotemark{1}} &
\colhead{$J$\tablenotemark{2}} &
\colhead{$H$\tablenotemark{2}} &
\colhead{$K$\tablenotemark{2}} &
\colhead{W1\tablenotemark{3}} &
\colhead{W2\tablenotemark{3}} &
\colhead{W3\tablenotemark{3}}
}
\startdata
201208431 & $16.23 \pm 0.05$ & $14.91 \pm 0.03$ & $15.56 \pm 0.04$ & $14.29 \pm 0.07$ & $13.89 \pm 0.12$ & $12.37 \pm 0.02$ & $11.75 \pm 0.02$ & $11.57 \pm 0.02$ & $11.51 \pm 0.02$ & $11.55 \pm 0.02$ & $11.58 \pm 0.20$   \\
201257461 & $12.82 \pm 0.03$ & $11.77 \pm 0.01$ & $12.24 \pm 0.04$ & $11.49 \pm 0.01$ & $11.19 \pm 0.02$ & $9.99 \pm 0.02$ & $9.48 \pm 0.02$ & $9.37 \pm 0.02$ & $9.28 \pm 0.02$ & $9.37 \pm 0.02$ & $9.30 \pm 0.04$ \\
201295312 & $12.78 \pm 0.04$ & $12.19 \pm 0.12$ & $12.41 \pm 0.03$ & $12.08 \pm 0.09$ & $12.01 \pm 0.21$ & $11.02 \pm 0.03$ & $10.70 \pm 0.02$ & $10.69 \pm 0.02$ & $10.63 \pm 0.02$ & $10.69 \pm 0.02$ & $10.75 \pm 0.12$ \\
201338508 & $16.30 \pm 0.07$ & $14.91 \pm 0.03$ & $15.62 \pm 0.05$ & $14.33 \pm 0.02$ & $13.79 \pm 0.05$ & $12.45 \pm 0.03$ & $11.76 \pm 0.02$ & $11.60 \pm 0.02$ & $11.49 \pm 0.03$ & $11.49 \pm 0.02$ & $11.16 \pm 0.13$ \\
201367065 & $13.52 \pm 0.06$ & $12.17 \pm 0.01$ & $12.87 \pm 0.03$ & $11.58 \pm 0.02$ & $10.98 \pm 0.17$ & $9.42 \pm 0.03$ & $8.80 \pm 0.04$ & $8.56 \pm 0.02$ & $8.44 \pm 0.02$ & $8.42 \pm 0.02$ & $8.32 \pm 0.02$ \\
201384232 & $13.30 \pm 0.05$ & $12.65 \pm 0.04$ & $12.91 \pm 0.05$ & $12.48 \pm 0.06$ & $12.34 \pm 0.07$ & $11.44 \pm 0.02$ & $11.09 \pm 0.02$ & $11.07 \pm 0.02$ & $11.00 \pm 0.02$ & $11.05 \pm 0.02$ & $11.21 \pm 0.16$ \\
201393098 & $13.90 \pm 0.04$ & $13.21 \pm 0.03$ & $13.54 \pm 0.06$ & $13.02 \pm 0.04$ & $12.85 \pm 0.05$ & $11.95 \pm 0.02$ & $11.63 \pm 0.02$ & $11.56 \pm 0.02$ & $11.52 \pm 0.02$ & $11.57 \pm 0.02$ & $11.61 \pm 0.21$ \\
201403446 & $12.48 \pm 0.02$ & $12.03 \pm 0.02$ & $12.18 \pm 0.01$ & $11.94 \pm 0.05$ & $11.86 \pm 0.04$ & $11.05 \pm 0.03$ & $10.76 \pm 0.02$ & $10.78 \pm 0.02$ & $10.67 \pm 0.03$ & $10.71 \pm 0.02$ & $10.36 \pm 0.07$ \\
201445392 & $15.73 \pm 0.02$ & $14.61 \pm 0.03$ & $15.19 \pm 0.04$ & $14.29 \pm 0.02$ & $14.03 \pm 0.07$ & $12.83 \pm 0.03$ & $12.32 \pm 0.03$ & $12.24 \pm 0.03$ & $12.16 \pm 0.02$ & $12.21 \pm 0.02$ & --- \\
201465501 & --- & --- & $16.73 \pm 0.02$ & $15.18 \pm 0.03$ & $14.35 \pm 0.15$ & $12.45 \pm 0.02$ & $11.71 \pm 0.02$ & $11.49 \pm 0.02$ & $11.35 \pm 0.02$ & $11.21 \pm 0.02$ & $11.35 \pm 0.19$ \\
201505350 & $13.80 \pm 0.02$ & $13.00 \pm 0.01$ & $13.36 \pm 0.02$ & $12.76 \pm 0.01$ & $12.57 \pm 0.02$ & $11.60 \pm 0.02$ & $11.21 \pm 0.02$ & $11.16 \pm 0.03$ & $11.10 \pm 0.02$ & $11.13 \pm 0.02$ & $10.95 \pm 0.12$ \\
201546283 & $13.51 \pm 0.07$ & $12.64 \pm 0.02$ & $13.03 \pm 0.02$ & $12.37 \pm 0.02$ & $12.17 \pm 0.05$ & $11.16 \pm 0.02$ & $10.79 \pm 0.03$ & $10.70 \pm 0.02$ & $10.61 \pm 0.02$ & $10.66 \pm 0.02$ & $10.53 \pm 0.09$ \\
201549860 & $15.56 \pm 0.06$ & $14.37 \pm 0.05$ & $14.95 \pm 0.07$ & $13.85 \pm 0.03$ & $13.45 \pm 0.05$ & $12.14 \pm 0.02$ & $11.56 \pm 0.02$ & $11.42 \pm 0.02$ & $11.38 \pm 0.02$ & $11.46 \pm 0.02$ & $11.60 \pm 0.25$ \\
201555883 & $16.48 \pm 0.01$ & $15.43 \pm 0.01$ & $16.19 \pm 0.10$ & $15.09 \pm 0.13$ & $14.55 \pm 0.08$ & $13.20 \pm 0.02$ & $12.53 \pm 0.03$ & $12.43 \pm 0.03$ & $12.34 \pm 0.02$ & $12.38 \pm 0.03$ & --- \\
201565013 & --- & --- & $18.25 \pm 0.01$ & $16.91 \pm 0.01$ & $16.34 \pm 0.01$ & $14.78 \pm 0.04$ & $14.11 \pm 0.05$ & $14.08 \pm 0.07$ & $13.94 \pm 0.03$ & $13.87 \pm 0.04$ & --- \\
201569483 & $12.90 \pm 0.08$ & $12.05 \pm 0.07$ & $12.44 \pm 0.03$ & $11.76 \pm 0.08$ & $11.48 \pm 0.08$ & $10.39 \pm 0.02$ & $9.97 \pm 0.03$ & $9.88 \pm 0.02$ & $9.82 \pm 0.02$ & $9.87 \pm 0.02$ & $9.82 \pm 0.05$ \\
201577035 & $13.14 \pm 0.11$ & $12.42 \pm 0.02$ & $12.70 \pm 0.04$ & $12.21 \pm 0.03$ & $12.13 \pm 0.20$ & $11.06 \pm 0.02$ & $10.75 \pm 0.02$ & $10.64 \pm 0.02$ & $10.64 \pm 0.02$ & $10.69 \pm 0.02$ & $10.55 \pm 0.10$ \\
201596316 & $14.21 \pm 0.01$ & $13.39 \pm 0.09$ & $13.78 \pm 0.07$ & $13.14 \pm 0.12$ & $12.88 \pm 0.10$ & $11.87 \pm 0.02$ & $11.46 \pm 0.02$ & $11.35 \pm 0.02$ & $11.29 \pm 0.02$ & $11.35 \pm 0.02$ & $10.80 \pm 0.11$ \\
201613023 & $12.99 \pm 0.09$ & $12.26 \pm 0.01$ & $12.56 \pm 0.03$ & $12.05 \pm 0.03$ & $11.96 \pm 0.08$ & $10.98 \pm 0.02$ & $10.71 \pm 0.02$ & $10.61 \pm 0.02$ & $10.58 \pm 0.02$ & $10.63 \pm 0.02$ & $10.59 \pm 0.10$ \\
201617985 & $16.34 \pm 0.02$ & $14.86 \pm 0.05$ & $15.62 \pm 0.06$ & $14.26 \pm 0.08$ & $13.42 \pm 0.09$ & $11.72 \pm 0.02$ & $11.09 \pm 0.04$ & $10.90 \pm 0.02$ & $10.73 \pm 0.02$ & $10.70 \pm 0.02$ & $10.86 \pm 0.11$ \\
201629650 & $13.61 \pm 0.03$ & $12.90 \pm 0.04$ & $13.20 \pm 0.03$ & $12.73 \pm 0.01$ & $12.53 \pm 0.06$ & $11.57 \pm 0.03$ & $11.26 \pm 0.02$ & $11.17 \pm 0.03$ & $11.14 \pm 0.02$ & $11.18 \pm 0.02$ & $10.93 \pm 0.12$ \\
201635569 & $17.74 \pm 0.16$ & $16.31 \pm 0.01$ & $17.02 \pm 0.01$ & $15.62 \pm 0.01$ & $14.87 \pm 0.01$ & $13.42 \pm 0.03$ & $12.77 \pm 0.02$ & $12.61 \pm 0.03$ & $12.52 \pm 0.03$ & $12.55 \pm 0.03$ & --- \\
201649426 & $14.57 \pm 0.03$ & $13.53 \pm 0.01$ & $14.04 \pm 0.01$ & $13.18 \pm 0.02$ & $12.86 \pm 0.06$ & $11.57 \pm 0.02$ & $11.07 \pm 0.02$ & $11.07 \pm 0.02$ & $10.88 \pm 0.02$ & $10.91 \pm 0.02$ & $10.86 \pm 0.12$ \\
201702477 & $15.27 \pm 0.05$ & $14.57 \pm 0.04$ & $14.89 \pm 0.04$ & $14.40 \pm 0.06$ & $14.24 \pm 0.03$ & $13.27 \pm 0.03$ & $12.88 \pm 0.03$ & $12.77 \pm 0.03$ & $12.81 \pm 0.02$ & $12.84 \pm 0.03$ & --- \\
201736247 & $15.49 \pm 0.06$ & $14.66 \pm 0.05$ & $15.01 \pm 0.04$ & $14.35 \pm 0.04$ & $14.14 \pm 0.02$ & $13.07 \pm 0.02$ & $12.55 \pm 0.02$ & $12.49 \pm 0.03$ & $12.46 \pm 0.02$ & $12.50 \pm 0.02$ & --- \\
201754305 & $15.65 \pm 0.04$ & $14.65 \pm 0.01$ & $15.13 \pm 0.04$ & $14.28 \pm 0.01$ & $13.93 \pm 0.05$ & $12.76 \pm 0.03$ & $12.21 \pm 0.03$ & $12.09 \pm 0.02$ & $12.06 \pm 0.02$ & $12.10 \pm 0.02$ & $12.34 \pm 0.46$ \\
201779067 & $11.81 \pm 0.01$ & $11.27 \pm 0.01$ & $11.53 \pm 0.07$ & $11.12 \pm 0.01$ & $10.95 \pm 0.01$ & $10.13 \pm 0.02$ & $9.87 \pm 0.02$ & $9.80 \pm 0.02$ & $9.74 \pm 0.02$ & $9.77 \pm 0.02$ & $9.74 \pm 0.04$ \\
201828749 & $12.48 \pm 0.04$ & $11.76 \pm 0.01$ & $12.13 \pm 0.05$ & $11.58 \pm 0.04$ & $11.32 \pm 0.04$ & $10.49 \pm 0.03$ & $10.23 \pm 0.04$ & $9.93 \pm 0.03$ & $9.82 \pm 0.02$ & $9.87 \pm 0.02$ & $9.98 \pm 0.06$ \\
201855371 & $14.82 \pm 0.06$ & $13.52 \pm 0.04$ & $14.20 \pm 0.06$ & $12.96 \pm 0.03$ & $12.45 \pm 0.01$ & $11.08 \pm 0.02$ & $10.44 \pm 0.02$ & $10.31 \pm 0.02$ & $10.22 \pm 0.02$ & $10.26 \pm 0.02$ & $10.12 \pm 0.07$ \\
201912552 & $15.01 \pm 0.06$ & $13.50 \pm 0.05$ & $14.22 \pm 0.05$ & $12.86 \pm 0.04$ & $11.66 \pm 0.08$ & $9.76 \pm 0.03$ & $9.13 \pm 0.03$ & $8.90 \pm 0.02$ & $8.77 \pm 0.02$ & $8.67 \pm 0.02$ & $8.55 \pm 0.03$ \\
201929294 & $14.32 \pm 0.04$ & $13.31 \pm 0.03$ & $13.78 \pm 0.05$ & $12.97 \pm 0.07$ & $12.61 \pm 0.09$ & $11.48 \pm 0.03$ & $10.98 \pm 0.02$ & $10.80 \pm 0.02$ & $10.73 \pm 0.02$ & $10.78 \pm 0.02$ & $10.67 \pm 0.10$
\enddata
\tablecomments{These data are available in interactive form at https://filtergraph.com/k2\_planets\_montet.}
\tablenotetext{1}{Magnitude from the AAVSO Photometric All-Sky Survey (APASS) DR6 \citep{Henden14} 
as reported in the UCAC4 Catalogue \citep{Zacharias12}.}
\tablenotetext{2}{Magnitude from the 2MASS All-Sky Catalog of Point Sources \citep{Cutri03}.}
\tablenotetext{3}{Magnitude from the ALLWise Data Release \citep{Cutri13}.}
\end{deluxetable*}
\end{turnpage}

%% file: table_stellarprops.tex
\clearpage
\begin{deluxetable*}{lcccccccc}
\tablewidth{0pt}
\tabletypesize{\scriptsize}
\tablecaption{Stellar Properties for all Objects of Interest \label{Tab:Stars}}
\tablehead{
\colhead{EPIC} &
\colhead{Alternate} &
\colhead{RA (J2000)} &
\colhead{Dec (J2000)} &
\colhead{Mass} &
\colhead{Radius} &
\colhead{\teff} &
\colhead{[Fe/H]} &
\colhead{Distance} \\
\colhead{} &
\colhead{Name} &
\colhead{(Degrees)} &
\colhead{(Degrees)} &
\colhead{($M_\odot$)} &
\colhead{($R_\odot$)} &
\colhead{(K)} &
\colhead{(dex)} &
\colhead{(pc)}
}
\startdata
 201208431 & K2-4 & $174.745639$ & $-3.905585$ & $0.63^{+0.03}_{-0.03}$ & $0.60^{+0.02}_{-0.02}$ & $4197^{+  45}_{ -43}$ & $-0.12^{+0.10}_{-0.12}$ & $ 218^{+  11}_{ -10}$ \\ 
 201257461 & ---     & $178.161110$ & $-3.094936$ & $1.50^{+0.04}_{-0.02}$ & $10.96^{+0.82}_{-0.93}$ & $5141^{+  38}_{ -42}$ & $-0.21^{+0.01}_{-0.01}$ & $1651^{+ 121}_{-134}$ \\ 
 201295312 & ---     & $174.011629$ & $-2.520881$ & $1.07^{+0.07}_{-0.07}$ & $1.09^{+0.20}_{-0.11}$ & $5989^{+ 100}_{ -81}$ & $-0.02^{+0.15}_{-0.18}$ & $ 331^{+  61}_{ -35}$ \\ 
 201338508 & K2-5 & $169.303502$ & $-1.877976$ & $0.53^{+0.01}_{-0.01}$ & $0.52^{+0.01}_{-0.01}$ & $4102^{+  45}_{ -41}$ & $-0.51^{+0.04}_{-0.06}$ & $ 181^{+   7}_{  -7}$ \\ 
 201367065 & K2-3 & $172.334949$ & $-1.454787$ & $0.53^{+0.02}_{-0.02}$ & $0.52^{+0.02}_{-0.02}$ & $3951^{+  33}_{ -38}$ & $-0.30^{+0.07}_{-0.06}$ & $  42^{+   2}_{  -2}$ \\ 
 201384232 & K2-6 & $178.192260$ & $-1.198477$ & $0.97^{+0.07}_{-0.07}$ & $0.96^{+0.14}_{-0.09}$ & $5850^{+  79}_{ -98}$ & $-0.14^{+0.17}_{-0.20}$ & $ 343^{+  52}_{ -33}$ \\ 
 201393098 & K2-7 & $167.093771$ & $-1.065755$ & $0.97^{+0.06}_{-0.06}$ & $0.96^{+0.17}_{-0.08}$ & $5772^{+  72}_{ -91}$ & $-0.07^{+0.16}_{-0.16}$ & $ 433^{+  75}_{ -38}$ \\ 
 201403446 & ---     & $174.266345$ & $-0.907261$ & $1.01^{+0.08}_{-0.06}$ & $1.12^{+0.26}_{-0.14}$ & $6445^{+  81}_{-111}$ & $-0.50^{+0.15}_{-0.13}$ & $ 362^{+  86}_{ -48}$ \\ 
 201445392 & K2-8 & $169.793666$ & $-0.284375$ & $0.79^{+0.03}_{-0.04}$ & $0.74^{+0.02}_{-0.03}$ & $4890^{+  38}_{ -58}$ & $-0.01^{+0.11}_{-0.13}$ & $ 405^{+  14}_{ -16}$ \\ 
 201465501 & K2-9 & $176.264467$ & $0.005301$ & $0.24^{+0.05}_{-0.03}$ & $0.25^{+0.04}_{-0.03}$ & $3468^{+  20}_{ -19}$ & $-0.46^{+0.12}_{-0.10}$ & $  66^{+  11}_{  -7}$ \\ 
 201505350 & K2-19& $174.960319$ & $0.603575$ & $0.84^{+0.04}_{-0.04}$ & $0.81^{+0.09}_{-0.05}$ & $5519^{+  49}_{ -82}$ & $-0.27^{+0.10}_{-0.10}$ & $ 291^{+  33}_{ -20}$ \\ 
 201546283 & ---     & $171.515164$ & $1.230738$ & $0.89^{+1.15}_{-0.07}$ & $0.88^{+7.37}_{-0.10}$ & $5422^{+ 194}_{ -93}$ & $-0.09^{+0.31}_{-0.15}$ & $ 251^{+2138}_{ -29}$ \\ 
 201549860 & ---     & $170.103081$ & $1.285956$ & $0.73^{+0.03}_{-0.03}$ & $0.69^{+0.02}_{-0.02}$ & $4523^{+  43}_{ -47}$ & $0.05^{+0.15}_{-0.14}$ & $ 249^{+   9}_{  -9}$ \\ 
 201555883 & ---     & $176.075940$ & $1.375947$ & $0.54^{+0.07}_{-0.01}$ & $0.52^{+0.08}_{-0.01}$ & $4419^{+  29}_{ -33}$ & $-0.98^{+0.62}_{-0.11}$ & $ 289^{+  46}_{  -9}$ \\ 
 201565013 & ---     & $176.992193$ & $1.510249$ & $0.51^{+0.13}_{-0.03}$ & $0.50^{+0.12}_{-0.03}$ & $3987^{+ 142}_{ -68}$ & $-0.44^{+0.47}_{-0.08}$ & $ 506^{+ 154}_{ -38}$ \\ 
 201569483 & ---     & $167.171300$ & $1.577513$ & $0.83^{+0.05}_{-0.05}$ & $0.79^{+0.06}_{-0.05}$ & $5192^{+  55}_{ -70}$ & $-0.09^{+0.17}_{-0.15}$ & $ 152^{+  12}_{ -10}$ \\ 
 201577035 & K2-10 & $172.121957$ & $1.690636$ & $0.94^{+0.04}_{-0.06}$ & $0.93^{+0.16}_{-0.07}$ & $5647^{+  60}_{ -89}$ & $-0.04^{+0.14}_{-0.17}$ & $ 271^{+  48}_{ -21}$ \\ 
 201596316 & K2-11 & $169.042002$ & $1.986840$ & $1.35^{+0.04}_{-0.56}$ & $5.15^{+0.20}_{-4.39}$ & $5433^{+  49}_{-144}$ & $-0.12^{+0.01}_{-0.17}$ & $2019^{+  71}_{-1728}$ \\ 
 201613023 & K2-12 & $173.192036$ & $2.244884$ & $1.01^{+0.05}_{-0.06}$ & $1.01^{+0.27}_{-0.09}$ & $5800^{+  53}_{ -90}$ & $0.03^{+0.13}_{-0.17}$ & $ 294^{+  78}_{ -27}$ \\ 
 201617985 & ---       & $179.491659$ & $2.321476$ & $0.52^{+0.03}_{-0.03}$ & $0.49^{+0.03}_{-0.03}$ & $3742^{+  31}_{ -36}$ & $-0.08^{+0.10}_{-0.11}$ & $ 111^{+   8}_{  -9}$ \\ 
 201629650 & K2-13 & $170.155529$ & $2.502696$ & $0.80^{+0.04}_{-0.04}$ & $0.78^{+0.09}_{-0.05}$ & $5698^{+  45}_{ -82}$ & $-0.54^{+0.12}_{-0.14}$ & $ 290^{+  34}_{ -18}$ \\ 
 201635569 & K2-14 & $178.057026$ & $2.594245$ & $0.47^{+0.01}_{-0.01}$ & $0.45^{+0.01}_{-0.01}$ & $3789^{+  17}_{ -16}$ & $-0.37^{+0.03}_{-0.04}$ & $ 219^{+   8}_{  -8}$ \\ 
 201649426 & ---       & $177.234262$ & $2.807619$ & $1.29^{+0.02}_{-0.02}$ & $8.15^{+0.32}_{-0.23}$ & $5086^{+  24}_{ -26}$ & $-0.17^{+0.01}_{-0.01}$ & $2537^{+  92}_{ -68}$ \\ 
 201702477 & ---       & $175.240794$ & $3.681584$ & $0.87^{+0.06}_{-0.06}$ & $0.85^{+0.11}_{-0.08}$ & $5618^{+  86}_{ -85}$ & $-0.26^{+0.17}_{-0.18}$ & $ 673^{+  87}_{ -63}$ \\ 
 201736247 & K2-15 & $178.110796$ & $4.254747$ & $0.72^{+0.06}_{-0.03}$ & $0.68^{+0.06}_{-0.03}$ & $5131^{+  69}_{ -65}$ & $-0.46^{+0.20}_{-0.14}$ & $ 437^{+  43}_{ -22}$ \\ 
 201754305 & K2-16 & $175.097258$ & $4.557340$ & $0.67^{+0.04}_{-0.03}$ & $0.64^{+0.03}_{-0.03}$ & $4761^{+  50}_{ -57}$ & $-0.40^{+0.12}_{-0.17}$ & $ 324^{+  16}_{ -16}$ \\ 
 201779067 & ---       & $168.542699$ & $4.988131$ & $0.91^{+0.03}_{-0.04}$ & $0.92^{+0.20}_{-0.07}$ & $6166^{+  30}_{ -51}$ & $-0.54^{+0.07}_{-0.12}$ & $ 188^{+  39}_{ -15}$ \\ 
 201828749 & ---       & $175.654343$ & $5.894323$ & $0.74^{+1.06}_{-0.04}$ & $0.71^{+9.64}_{-0.06}$ & $5552^{+  87}_{ -97}$ & $-0.69^{+0.34}_{-0.23}$ & $ 146^{+1996}_{ -12}$ \\ 
 201855371 & K2-17 & $178.329776$ & $6.412261$ & $0.71^{+0.02}_{-0.05}$ & $0.66^{+0.02}_{-0.03}$ & $4320^{+  56}_{ -47}$ & $0.15^{+0.09}_{-0.22}$ & $ 134^{+   5}_{  -6}$ \\ 
 201912552\tablenotemark{1} & K2-18 & $172.560461$ & $7.588391$ & $0.413^{+0.043}_{-0.043}$ & $0.394^{+0.038}_{-0.038}$ & $3503^{+  60}_{ -60}$ & $0.09^{+0.09}_{-0.09}$ & $  34^{+   4}_{  -4}$ \\ 
 201929294 & ---       & $174.656968$ & $7.959611$ & $0.73^{+0.06}_{-0.09}$ & $0.70^{+0.04}_{-0.08}$ & $4786^{+  48}_{ -53}$ & $-0.16^{+0.22}_{-0.34}$ & $ 197^{+  13}_{ -24}$ \\ 
\enddata
\tablecomments{These values and their uncertainties are derived from \texttt{MULTINEST} analysis and the
numbers are computed as the 0.158, 0.500, and 0.842 posterior sample quantiles. 
The coordinates are retrieved directly from the EPIC. These data are available in interactive form at https://filtergraph.com/k2\_planets\_montet.}
\tablenotetext{1}{Parameters inferred from spectroscopic observations.}
\end{deluxetable*}

%% file: table_planets.tex
\clearpage
\begin{deluxetable*}{lccccccc}
\tablewidth{0pt}
\tabletypesize{\scriptsize}
\tablecaption{Planet Properties for all Objects of Interest \label{Tab:Planets}}
\tablehead{
\colhead{Candidate} &
\colhead{Period (days)} &
\colhead{Epoch (BJD-2456808)} &
\colhead{Radius ($R_\oplus$)} &
\colhead{$a/R_\star$} &
\colhead{$a$ (AU)} &
\colhead{$T_\mathrm{eq}$ (K)} &
\colhead{Disposition}
}
\startdata
$201208431.01$/K2-4b & $10.00329 \pm {0.00159}$ & $7.5212 \pm {0.0080}$ & $2.37 \pm {0.40}$ & $27.79 \pm {0.72}$ & $0.0777 \pm {0.0012}$ & $563 \pm {11} $     & Planet\\
$201257461.01$ & $50.27762 \pm {0.00785}$ & $20.3735 \pm {0.0397}$ & $209.52 \pm {99.23}$ & $6.19 \pm {0.52}$ & $0.3049 \pm {0.0030}$ & $1466 \pm {52} $ & FP \\
$201295312.01$ & $5.65706 \pm {0.00079}$ & $3.7187 \pm {0.0082}$ & $2.16 \pm {0.57}$ & $12.94 \pm {4.07}$ & $0.0633 \pm {0.0019}$ & $1211 \pm {154} $    & Candidate \\
$201338508.01$/K2-5c & $10.93406 \pm {0.00205}$ & $6.5947 \pm {0.0080}$ & $1.92 \pm {0.20}$ & $32.27 \pm {0.71}$ & $0.0783 \pm {0.0007}$ & $511 \pm {9} $      & Planet \\
$201338508.02$/K2-5b & $5.73491 \pm {0.00061}$ & $0.8640 \pm {0.0063}$ & $1.92 \pm {0.23}$ & $20.99 \pm {0.46}$ & $0.0509 \pm {0.0004}$ & $634 \pm {12} $      & Planet \\
$201367065.01$/K2-3b & $10.05448 \pm {0.00033}$ & $5.4177 \pm {0.0015}$ & $1.98 \pm {0.10}$ & $30.72 \pm {0.75}$ & $0.0740 \pm {0.0009}$ & $504 \pm {9} $      & Planet \\
$201367065.02$/K2-3c & $24.64745 \pm {0.00152}$ & $4.2759 \pm {0.0030}$ & $1.56 \pm {0.10}$ & $55.85 \pm {1.36}$ & $0.1345 \pm {0.0016}$ & $374 \pm {7} $      & Planet \\
$201384232.01$/K2-6b & $30.94191 \pm {0.00467}$ & $19.5014 \pm {0.0090}$ & $2.50 \pm {0.88}$ & $50.27 \pm {24.56}$ & $0.1898 \pm {0.0056}$ & $615 \pm {105} $  & Planet \\
$201393098.01$/K2-7b & $28.67992 \pm {0.00947}$ & $16.6155 \pm {0.0149}$ & $2.67 \pm {0.56}$ & $40.29 \pm {8.19}$ & $0.1814 \pm {0.0043}$ & $651 \pm {61} $    & Planet \\
$201403446.01$ & $19.15344 \pm {0.00607}$ & $7.3412 \pm {0.0152}$ & $2.04 \pm {0.46}$ & $27.05 \pm {5.87}$ & $0.1408 \pm {0.0040}$ & $889 \pm {88} $     & Candidate \\
$201445392.01$/K2-8b & $10.35176 \pm {0.00133}$ & $5.6119 \pm {0.0053}$ & $2.97 \pm {0.51}$ & $24.94 \pm {0.79}$ & $0.0856 \pm {0.0012}$ & $691 \pm {14} $     & Planet \\
$201445392.02$ & $5.06468 \pm {0.00063}$ & $5.0663 \pm {0.0071}$ & $2.31 \pm {0.33}$ & $15.49 \pm {0.49}$ & $0.0531 \pm {0.0008}$ & $877 \pm {17} $      & Candidate \\
$201465501.01$/K2-9b & $18.44883 \pm {0.00137}$ & $14.6723 \pm {0.0030}$ & $1.60 \pm {0.42}$ & $74.76 \pm {6.66}$ & $0.0848 \pm {0.0050}$ & $284 \pm {14} $    & Planet \\
$201505350.01$/K2-19c & $11.90691 \pm {0.00037}$ & $9.2764 \pm {0.0018}$ & $4.31 \pm {0.49}$ & $24.09 \pm {2.48}$ & $0.0965 \pm {0.0017}$ & $797 \pm {42} $     & Planet \\
$201505350.02$/K2-19b & $7.91943 \pm {0.00007}$ & $5.3836 \pm {0.0005}$ & $7.11 \pm {0.81}$ & $18.35 \pm {1.89}$ & $0.0735 \pm {0.0013}$ & $913 \pm {48} $      & Planet \\
$201546283.01$ & $6.77131 \pm {0.00012}$ & $4.8440 \pm {0.0022}$ & $5.77 \pm {3.24}$ & $17.56 \pm {9.24}$ & $0.0668 \pm {0.0029}$ & $991 \pm {239} $     & Candidate \\
$201549860.01$ & $5.60840 \pm {0.00055}$ & $4.1181 \pm {0.0047}$ & $2.20 \pm {0.40}$ & $17.42 \pm {0.46}$ & $0.0555 \pm {0.0008}$ & $766 \pm {14} $      & Candidate \\
$201555883.01$ & --- & --- & --- & --- & --- & --- & FP\tablenotemark{2}\\
$201565013.01$ & $8.63810 \pm {0.00024}$ & $3.4284 \pm {0.0016}$ & $15.99 \pm {9.19}$ & $28.07 \pm {2.68}$ & $0.0669 \pm {0.0031}$ & $536 \pm {37} $     & Candidate \\
$201569483.01$ & $5.79687 \pm {0.00000}$ & $5.3135 \pm {0.0004}$ & $27.81 \pm {3.56}$ & $15.68 \pm {1.91}$ & $0.0589 \pm {0.0015}$ & $930 \pm {51} $     & FP \\
$201577035.01$/K2-10b & $19.30691 \pm {0.00127}$ & $11.5768 \pm {0.0033}$ & $3.92 \pm {0.69}$ & $32.74 \pm {5.15}$ & $0.1374 \pm {0.0025}$ & $703 \pm {55} $    & Planet \\
$201596316.01$/K2-11b & $39.93767 \pm {0.23229}$ & $21.8290 \pm {0.1156}$ & $7.55 \pm {9.33}$ & $45.08 \pm {58.53}$ & $0.2257 \pm {0.0143}$ & $734 \pm {253} $  & Planet \\
$201613023.01$/K2-12b & $8.28212 \pm {0.00060}$ & $7.3734 \pm {0.0054}$ & $2.33 \pm {0.58}$ & $17.47 \pm {5.05}$ & $0.0802 \pm {0.0021}$ & $1003 \pm {121} $    & Planet \\
$201617985.01$ & $7.28161 \pm {0.00078}$ & $4.6366 \pm {0.0047}$ & $1.78 \pm {0.43}$ & $26.04 \pm {1.16}$ & $0.0586 \pm {0.0012}$ & $518 \pm {16} $      & Candidate \\
$201629650.01$/K2-13b & $39.91488 \pm {0.32477}$ & $4.5250 \pm {0.0146}$ & $1.89 \pm {0.95}$ & $79.69 \pm {63.37}$ & $0.2114 \pm {0.0061}$ & $511 \pm {126} $   & Planet \\
$201635569.01$/K2-14b & $8.36802 \pm {0.00019}$ & $3.4513 \pm {0.0013}$ & $4.81 \pm {0.42}$ & $30.16 \pm {0.69}$ & $0.0627 \pm {0.0006}$ & $488 \pm {8} $       & Planet \\
$201649426.01$ & $27.77045 \pm {0.00008}$ & $13.3482 \pm {0.0012}$ & $32.79 \pm {9.01}$ & $59.26 \pm {13.58}$ & $0.1517 \pm {0.0097}$ & $441 \pm {42} $  & FP \\
$201702477.01$ & $40.73620 \pm {0.00266}$ & $3.5455 \pm {0.0025}$ & $7.28 \pm {1.10}$ & $56.98 \pm {7.61}$ & $0.2205 \pm {0.0053}$ & $529 \pm {36} $     & Candidate \\
$201736247.01$/K2-15b & $11.81040 \pm {0.00204}$ & $3.8509 \pm {0.0076}$ & $2.48 \pm {0.30}$ & $28.84 \pm {1.98}$ & $0.0910 \pm {0.0018}$ & $676 \pm {26} $     & Planet \\
$201754305.01$/K2-16c & $19.07536 \pm {0.00490}$ & $1.4854 \pm {0.0119}$ & $2.14 \pm {0.41}$ & $41.43 \pm {1.34}$ & $0.1220 \pm {0.0021}$ & $523 \pm {12} $     & Planet \\
$201754305.02$/K2-16b & $7.62067 \pm {0.00095}$ & $3.6802 \pm {0.0054}$ & $2.13 \pm {0.37}$ & $22.47 \pm {0.73}$ & $0.0662 \pm {0.0011}$ & $710 \pm {16} $      & Planet \\
$201779067.01$ & $27.24273 \pm {0.00012}$ & $12.2601 \pm {0.0003}$ & $31.73 \pm {5.25}$ & $38.25 \pm {3.72}$ & $0.1718 \pm {0.0022}$ & $707 \pm {34} $   & FP \\
$201828749.01$ & $33.51569 \pm {0.00232}$ & $5.1504 \pm {0.0034}$ & $3.83 \pm {3.25}$ & $67.09 \pm {67.64}$ & $0.1875 \pm {0.0090}$ & $613 \pm {239} $   & Candidate \\
$201855371.01$/K2-17b & $17.96753 \pm {0.00152}$ & $9.9462 \pm {0.0035}$ & $2.23 \pm {0.20}$ & $39.38 \pm {0.85}$ & $0.1190 \pm {0.0020}$ & $487 \pm {10} $     & Planet \\
$201912552.01$/K2-18b\tablenotemark{1} & $32.94488 \pm {0.00281}$ & $28.1849 \pm {0.0027}$ & $2.24 \pm {0.23}$ & $83.83 \pm {9.03}$ & $0.1491 \pm {0.0055}$ & $272 \pm {15}$ & Planet \\
$201929294.01$ & --- & --- & --- & --- & --- & --- & FP\tablenotemark{2}\\
\enddata
\tablecomments{These values and uncertainties are given by the mean and
standard deviation of MCMC posterior samplings. These data are available in interactive form at https://filtergraph.com/k2\_planets\_montet.}
\tablenotetext{1}{Parameters inferred from spectroscopic observations.}
\tablenotetext{2}{Declared a false positive due to noise modeling systematics (see \textsection 5.2)}
\end{deluxetable*}

%% file: table_bg.tex
\clearpage
\begin{deluxetable*}{lcccccccc}
\tablewidth{0pt}
\tabletypesize{\scriptsize}
\tablecaption{Detected Companions to Candidate Host Stars \label{Tab:bg}}
\tablehead{
\colhead{Primary} &
\colhead{Aperture\tablenotemark{1}} &
\colhead{RA\tablenotemark{2}} &
\colhead{Dec\tablenotemark{2}} &
\colhead{Detection\tablenotemark{3}} &
\colhead{Separation\tablenotemark{4}} &
\colhead{$\Delta r$\tablenotemark{5}} &
\colhead{Max Depth\tablenotemark{6}} &
\colhead{Observed Depth\tablenotemark{7}} \\
\colhead{} &
\colhead{(arcsec)} &
\colhead{(J2000)} &
\colhead{(J2000)} &
\colhead{} &
\colhead{(arcsec)} &
\colhead{(mag)} &
\colhead{(ppt)} &
\colhead{(ppt)}
}
\startdata
 201208431  & 15.9 & 174.748988  & -3.902146  &  SDSS  &       $17.25 \pm 0.15\tablenotemark{b}$  & $5.90 \pm 0.12$ & 5.6 & 1.20 \\
 201257461  & 19.9 & 178.164376  & -3.093431  &  SDSS  &       $ 12.91 \pm 0.18\tablenotemark{b}$ & $5.04 \pm 0.03$ & 4.8 & 30.54 \\
 201295312  & 11.9 &  174.010158  & -2.522528  &  SDSS/AO &  $8.12 \pm 0.09$\tablenotemark{b}   & $7.10 \pm 0.10$ & 0.8 & 0.30 \\
 201338508  & 15.9 & 169.308176  & -1.873647  & SDSS      &    $22.92 \pm 0.07$\tablenotemark{b} & $4.35 \pm 0.03$ & 9.1 & 1.07 \\
 201367065  & 19.9 &                  &                       &                &                                                            &                          &          & 1.26  \\
 201384232  & 13.9 & 178.195303 &  -1.192501  & SDSS      &   $24.14 \pm 0.06$\tablenotemark{b} & $5.93 \pm 0.03$ & 2.1 & 0.68 \\
 201393098  & 15.9 &                  &                       &                &                                                           &                          &          & 0.53  \\
 201403446  & 15.9 & 174.267663 &  -0.909645  & SDSS      &    $9.78 \pm 0.14$\tablenotemark{b} & $4.56 \pm 0.08$ & 7.5 & 0.23 \\ 
 201445392  & 13.9 &                &                       &                &                                                           &                          &             & 0.78 \\
 201465501  & 11.9 &                 &                       &                &                                                           &                          &            & 2.83 \\
 201505350  & 19.9 &                  &                       &                &                                                           &                          &           & 2.64  \\
 201546283  & 17.9 & 171.515265  &  1.229950  &  SDSS/AO &  $2.98 \pm 0.05$\tablenotemark{a}   & $5.87 \pm 0.06$ & 2.3 & 2.33 \\
 201549860  & 13.9 & 170.097556 &  1.288007    & SDSS    &    $21.21 \pm 0.05$\tablenotemark{b} & $2.26 \pm 0.03$ & 62.3 & 0.80 \\
 201555883  & 10.0 &                 &                       &                &                                                           &                          &              & 3.50 \\
 201565013  & 10.0 &               &                       &                &                                                           &                          &                & 45.8    \\
 201569483  & 19.9 &                 &                       &                &                                                           &                          &              & 160  \\
 201577035  & 19.9 & 172.118116 &  1.687798    & SDSS    &  $17.19 \pm 0.12$\tablenotemark{b}  & $5.40 \pm 0.03$ & 3.5    & 1.44  \\
 201596316  & 15.9 &                 &                       &                &                                                           &                          &              & 0.70 \\
 201613023  & 19.9 &                 &                       &                &                                                           &                          &              & 0.42 \\
 201617985  & 15.9 &                 &                       &                &                                                           &                          &             & 1.10  \\
 201629650  & 15.9 & 170.158905 &  2.502107   & SDSS     & $12.30 \pm 0.14$\tablenotemark{b} &   $5.98 \pm 0.06$ & 2.0   & 0.58 \\
 201635569  & 11.9 &                  &                       &                &                                                           &                          &            & 9.43 \\
 201649426  & 19.9 &                 &                       &                &                                                           &                          &             & 216   \\
 201702477  & 10.0 & 175.238916 &  3.678764    & SDSS     &  $12.15 \pm 0.12$\tablenotemark{b} & $4.65 \pm 0.09$ & 6.9   & 6.70 \\
 201736247  & 13.9 &                  &                       &                &                                                           &                          &            & 1.21 \\
 201754305  & 11.9 &                  &                       &                &                                                           &                          &            & 0.80  \\
 201779067  & 19.9 &                  &                       &                &                                                           &                          &            & 84.9  \\
 201828749  & 11.9 & 175.645724  &  5.894714  &   AO  &        $2.46 \pm 0.04$\tablenotemark{a}   & $ 2.0 \pm 0.1 $\tablenotemark{c} & 137 & 0.76  \\
 201855371  & 19.9 &                  &                       &                &                                                           &                          &            & 0.99   \\
 201912552  & 13.9 &                 &                       &                &                                                           &                          &             & 2.85   \\
 201929294  & 19.9 &                  &                       &                &                                                           &                          &            & 13.56 
\enddata
\tablenotetext{1}{Defined aperture used to create the \KT\ stellar light curve.}
\tablenotetext{2}{Position of imaged companion.}
\tablenotetext{3}{Dataset used to detect the imaged companion.}
\tablenotetext{4}{Distance between the primary \KT\ target star and companion, in the dataset in which the
companion is detected.}
\tablenotetext{5}{Difference in $r$-band magnitude between the primary \KT\ target star and the companion.}
\tablenotetext{6}{Observed ``transit'' depth if the imaged companion's flux were fully contained in the aperture and
if it were an equal-mass eclipsing binary, leading to an eclipse depth of 50\%. This is the maximum possible false positive
eclipse depth, as described in \textsection \ref{sec:background}.}
\tablenotetext{7}{Observed transit depth in the \KT\ dataset. If larger than the "max depth," this transit event cannot
be caused by eclipses of the background star.}
\tablenotetext{a}{Separation from AO imaging}
\tablenotetext{b}{Separation from SDSS photometry}
\tablenotetext{c}{$\Delta r$ inferred from $JHK$ relative photometry.}
\end{deluxetable*}

%% file: table_fpp.tex
\clearpage
\begin{deluxetable*}{lcccccccc}
\tablewidth{0pt}
\tabletypesize{\scriptsize}
\tablecaption{False Positive Probability Calculation Results}
\label{Table:FPP}
\tablehead{
\colhead{Candidate} &
\colhead{$\delta_{\rm sec, max}$ [ppt]\tablenotemark{1}} &
\colhead{AO?\tablenotemark{2}} &
\colhead{${\rm Pr}_{\rm EB}$} &
\colhead{${\rm Pr}_{\rm BEB}$} &
\colhead{${\rm Pr}_{\rm HEB}$} &
\colhead{$f_{p}$\tablenotemark{3}} &
\colhead{FPP} &
\colhead{Disposition}
}
\startdata
201208431.01/K2-4b & $0.51$ &  - & $< 10^{-4}$ & $8.1\times10^{-4}$ & $< 10^{-4}$ & $0.21$ & $8.1\times10^{-4}$ & Planet  \\
 \color{red} 201257461.01  & \color{red}  $0.59$  & \color{red}   -  & \color{red}  $0.998$  & \color{red}  $1.7\times10^{-3}$  & \color{red}  $< 10^{-4}$  & \color{red}  $0.00$  & \color{red}  $1.000$  & \color{red}  FP \\
201295312.01 & $0.04$ &  Y & $1.4\times10^{-4}$ & $< 10^{-4}$ & $< 10^{-4}$ & $0.17$ & $1.4\times10^{-4}$ & Candidate\tablenotemark{a}  \\
201338508.01/K2-5c & $0.63$ &  - & $< 10^{-4}$ & $2.9\times10^{-3}$ & $< 10^{-4}$ & $0.22$ & $2.9\times10^{-3}$ & Planet  \\
201338508.02/K2-5b & $0.33$ &  - & $< 10^{-4}$ & $1.7\times10^{-4}$ & $< 10^{-4}$ & $0.22$ & $1.7\times10^{-4}$ & Planet  \\
201367065.01/K2-3b & $0.15$ &  - & $< 10^{-4}$ & $1.1\times10^{-4}$ & $< 10^{-4}$ & $0.22$ & $1.1\times10^{-4}$ & Planet\tablenotemark{c}  \\
201367065.02/K2-3c & $0.67$ &  - & $< 10^{-4}$ & $< 10^{-4}$ & $< 10^{-4}$ & $0.16$ & $< 10^{-4}$ & Planet\tablenotemark{c}  \\
201384232.01/K2-6b & $0.44$ &  - & $8.4\times10^{-3}$ & $< 10^{-4}$ & $< 10^{-4}$ & $0.07$ & $8.5\times10^{-3}$ & Planet  \\
201393098.01/K2-7b & $0.52$ &  - & $< 10^{-4}$ & $1.1\times10^{-3}$ & $< 10^{-4}$ & $0.05$ & $1.1\times10^{-3}$ & Planet  \\
201403446.01 & $0.18$ &  Y & $4.8\times10^{-4}$ & $< 10^{-4}$ & $< 10^{-4}$ & $0.19$ & $4.9\times10^{-4}$ & Candidate\tablenotemark{a}  \\
201445392.01/K2-8b & $0.26$ &  - & $< 10^{-4}$ & $2.1\times10^{-3}$ & $< 10^{-4}$ & $0.18$ & $2.1\times10^{-3}$ & Planet  \\
201445392.02 & $0.18$ &  - & $< 10^{-4}$ & $0.019$ & $< 10^{-4}$ & $0.21$ & $0.019$ & Candidate  \\
201465501.01/K2-9b & $0.68$ &  - & $< 10^{-4}$ & $5.8\times10^{-3}$ & $< 10^{-4}$ & $0.21$ & $5.8\times10^{-3}$ & Planet  \\
201505350.01/K2-19c & $2.69$ &  - & $< 10^{-4}$ & $5.6\times10^{-3}$ & $< 10^{-4}$ & $0.04$ & $5.6\times10^{-3}$ & Planet\tablenotemark{d}  \\
201505350.02/K2-19b & $0.70$ &  - & $< 10^{-4}$ & $1.6\times10^{-4}$ & $< 10^{-4}$ & $0.07$ & $1.7\times10^{-4}$ & Planet\tablenotemark{d}  \\
201546283.01 & $0.15$ &  - & $7.0\times10^{-4}$ & $2.6\times10^{-4}$ & $< 10^{-4}$ & $0.00$ & $9.6\times10^{-4}$ & Candidate\tablenotemark{a}  \\
201549860.01 & $0.18$ &  - & $< 10^{-4}$ & $0.026$ & $< 10^{-4}$ & $0.04$ & $0.026$ & Candidate  \\
 \color{red} 201555883.01  & \color{red}  $0.94$  & \color{red}   -  & \color{red}   --  & \color{red}   --  & \color{red}   --  & \color{red}   --  & \color{red}  --  & \color{red}  FP\tablenotemark{b} \\
201565013.01 & $1.69$ &  - & $0.783$ & $7.3\times10^{-3}$ & $0.063$ & $0.07$ & $0.853$ & Candidate  \\
 \color{red} 201569483.01  & \color{red}  $2.06$  & \color{red}   -  & \color{red}  $0.822$  & \color{red}  $< 10^{-4}$  & \color{red}  $0.174$  & \color{red}  $0.00$  & \color{red}  $0.996$  & \color{red}  FP \\
201577035.01/K2-10b & $0.14$ &  Y & $4.4\times10^{-4}$ & $< 10^{-4}$ & $< 10^{-4}$ & $0.07$ & $4.4\times10^{-4}$ & Planet  \\
201596316.01/K2-11b & $0.45$ &  - & $< 10^{-4}$ & $1.2\times10^{-3}$ & $< 10^{-4}$ & $0.06$ & $1.2\times10^{-3}$ & Planet  \\
201613023.01/K2-12b & $0.08$ &  Y & $< 10^{-4}$ & $< 10^{-4}$ & $< 10^{-4}$ & $0.18$ & $< 10^{-4}$ & Planet  \\
201617985.01 & $0.27$ &  - & $< 10^{-4}$ & $0.012$ & $< 10^{-4}$ & $0.18$ & $0.012$ & Candidate  \\
201629650.01/K2-13b & $0.43$ &  - & $5.9\times10^{-4}$ & $2.0\times10^{-4}$ & $< 10^{-4}$ & $0.13$ & $7.8\times10^{-4}$ & Planet  \\
201635569.01/K2-14b & $0.79$ &  - & $< 10^{-4}$ & $4.9\times10^{-3}$ & $< 10^{-4}$ & $0.05$ & $4.9\times10^{-3}$ & Planet  \\
 \color{red} 201649426.01  & \color{red}  $3.10$  & \color{red}   -  & \color{red}  $0.896$  & \color{red}  $< 10^{-4}$  & \color{red}  $0.104$  & \color{red}  $0.00$  & \color{red}  $1.000$  & \color{red}  FP \\
201702477.01 & $0.70$ &  - & $0.137$ & $1.2\times10^{-3}$ & $6.6\times10^{-3}$ & $0.05$ & $0.145$ & Candidate  \\
201736247.01/K2-15b & $0.42$ &  - & $4.8\times10^{-4}$ & $2.1\times10^{-4}$ & $< 10^{-4}$ & $0.19$ & $6.9\times10^{-4}$ & Planet  \\
201754305.01/K2-16c & $0.65$ &  - & $1.0\times10^{-4}$ & $1.4\times10^{-3}$ & $< 10^{-4}$ & $0.21$ & $1.5\times10^{-3}$ & Planet  \\
201754305.02/K2-16b & $0.38$ &  - & $2.3\times10^{-4}$ & $9.9\times10^{-4}$ & $< 10^{-4}$ & $0.19$ & $1.2\times10^{-3}$ & Planet  \\
 \color{red} 201779067.01  & \color{red}  $1.97$  & \color{red}   -  & \color{red}  $0.968$  & \color{red}  $1.3\times10^{-3}$  & \color{red}  $7.2\times10^{-3}$  & \color{red}  $0.00$  & \color{red}  $0.976$  & \color{red}  FP \\
201828749.01 & $0.39$ &  Y & $0.644$ & $3.8\times10^{-4}$ & $< 10^{-4}$ & $0.01$ & $0.645$ & Candidate  \\
201855371.01/K2-17b & $0.62$ &  - & $< 10^{-4}$ & $8.7\times10^{-3}$ & $< 10^{-4}$ & $0.01$ & $8.7\times10^{-3}$ & Planet  \\
201912552.01/K2-18b & $0.47$ &  Y & $< 10^{-4}$ & $< 10^{-4}$ & $< 10^{-4}$ & $0.21$ & $< 10^{-4}$ & Planet  \\
 \color{red} 201929294.01  & \color{red}  $3.12$  & \color{red}   -  & \color{red}   --  & \color{red}   --  & \color{red}   --  & \color{red}   --  & \color{red}  --  & \color{red}  FP\tablenotemark{b}  
\enddata
\tablecomments{Results of the \texttt{vespa} astrophysical 
false positive probability calculations for all candidates.  
Likely false positives (FPP $> 0.9$, or otherwise designated)
 are marked in red.  
Candidates are declared to be validated planets if FPP $< 0.01$.  
EB, BEB, and HEB refer to the three considered astrophysical 
false positive scenarios, and the relative probability of 
each is listed in the appropriate column.  Planets previously 
identified in the literature are marked.}
\tablenotetext{1}{Maximum depth of potential secondary eclipse signal.}
\tablenotetext{2}{Whether adaptive optics observation is presented in this paper.}
\tablenotetext{3}{Integrated planet occurrence rate assumed between 0.7$\times$ and 1.3$\times$ the candidate radius}
\tablenotetext{a}{Despite low FPP, returned to candidate status 
out of abundance of caution due to secondary star detection within or near photometric aperture.}
\tablenotetext{b}{Declared a false positive due to noise modeling systematics (see \S\ref{sec:systematics}).}
\tablenotetext{c}{Identified as planets by \citet{Crossfield15}.}
\tablenotetext{d}{Identified as planets by \citet{Armstrong15b}.}
\end{deluxetable*}